\documentclass[10pt,letterpaper]{article}
\usepackage[top=0.85in,left=0.85in,footskip=0.75in]{geometry}

\usepackage{changepage}

\usepackage[utf8x]{inputenc}

\usepackage{textcomp,marvosym}

\usepackage{fixltx2e}

\usepackage{amsmath,amssymb}

\usepackage{cite}

\usepackage{nameref,hyperref}

\usepackage{microtype}
\DisableLigatures[f]{encoding = *, family = * }

\raggedright
\usepackage[aboveskip=1pt,labelfont=bf,labelsep=period,justification=raggedright,singlelinecheck=off]{caption}

\makeatletter
\renewcommand{\@biblabel}[1]{\quad#1.}
\makeatother

\usepackage{lastpage,fancyhdr,graphicx}
\usepackage{epstopdf}

\newcommand{\sqb}[1]{\left[ #1 \right]}
\newcommand{\rndb}[1]{\left( #1 \right)}

\newcommand{\norm}[1]{\left| #1 \right|}
\newcommand{\vect}[1]{\boldsymbol{#1}}
\newcommand{\tens}[1]{\underline{\underline{#1}}}
\newcommand{\diff}[2]{\frac{{\rm d}#1}{{\rm d}#2}}
\newcommand{\pdiff}[2]{\frac{\partial#1}{\partial#2}}
\newcommand{\divv}{\nabla \cdot}

\newcommand{\act}{\zeta \Delta \mu}
\newcommand{\actc}{\zeta_c \Delta \mu}
\newcommand{\actt}{\tilde{\zeta} \Delta \mu}

\begin{document}
\vspace*{0.2in}

\begin{flushleft}
{\Large
\textbf\newline{Immersed Boundary Simulations of Active Fluid Droplets} 
}
\newline
\\
Carl A. Whitfield\textsuperscript{1 \textcurrency *} and
Rhoda J. Hawkins\textsuperscript{1}.
\\
\bigskip
\textbf{1} Department of Physics and Astronomy, University of Sheffield, Sheffield, UK, S3 7RH
\\
\bigskip

\textsuperscript{\textcurrency}Current Address: Department of Physics, University of Warwick, Coventry, UK, CV3 7AL

*Corresponding author: carl.whitfield@physics.org

\end{flushleft}
\section*{Abstract}
We present numerical simulations of active fluid droplets immersed in an external fluid in 2-dimensions { using} an Immersed Boundary method to simulate the fluid droplet interface as a Lagrangian mesh. We present results from two example systems, firstly an active isotropic fluid boundary consisting of particles that can bind and unbind from the interface and generate surface tension gradients through active contractility. Secondly, a droplet filled with an active polar fluid with { homeotropic} anchoring at the droplet interface. These two systems demonstrate spontaneous symmetry breaking and steady state dynamics resembling cell motility and division and show complex feedback mechanisms with minimal degrees of freedom. The simulations outlined here will be useful for quantifying the wide range of dynamics observable in these active systems and modelling the effects of confinement in a consistent and adaptable way.

\section*{Introduction}
\label{sec:intro6}

Active fluids are ubiquitous in biology and include the cell cytoskeleton \cite{Zumdieck2005,Joanny2009}, bacteria films and even schools of fish \cite{Ramaswamy2010,Marchetti2013}. The common trait of these systems is that they are driven out of equilibrium at the level of their constituents, which interact hydrodynamically with their neighbours. In this paper we use a continuum model of active fluids developed in \cite{Kruse2004,Kruse2005,Furthauer2012} where the constituent fluid particles generate local force dipoles. These systems demonstrate aspects of cytoskeletal dynamics in cells or collections of microswimmers such as bacteria, { which} generate dipolar stresses in their surrounding medium. Recent experimental work has shown that it is possible to confine these systems to droplets or vesicles \emph{in vitro} to isolate their dynamics and observe emergent behaviour \cite{Tsai2011, Sanchez2012, Keber2014, Vladescu2014a}. With these systems as motivation we consider two limits of the active fluid model, firstly the case of an isotropically ordered active fluid confined to a droplet interface and secondly the case of a droplet of active fluid with strong local polar ordering. We are able to capture the full dynamics of these two systems in a single computational model that predicts interesting phase behaviour in both cases.

We have developed numerical simulations to model active fluids confined within and on the interface of deformable droplets in 2-dimensions. The simulations are based on the Immersed Boundary method introduced by Peskin \cite{Peskin2002} to model interactions between fluids and elastic interfaces in biological systems. This method models the droplet interface as a 1D Lagrangian mesh of points which interacts with a fixed fluid mesh via a numerical analogue of the Dirac delta function (as outlined in the Methods section). The advantage of this method over phase field methods (such as \cite{Yue2004a, Tjhung2012, Giomi2014}) is that we can track and numerically conserve quantities defined on the interface naturally in the simulation and potentially model a wide ranges of interface properties. Moreover, we can easily incorporate external boundaries in a self-consistent manner. However, a limitation of this method is that it is significantly more difficult to consider droplet separation or combination, phenomena which are captured automatically by phase field methods.

{ Applying the immersed boundary method to active fluids is a natural step as the properties and geometries of biological interfaces can be complex. We simulate the coupled dynamics of active particles on the interface as well as in the droplet bulk in a self consistent way. Firstly, this allows us to exactly conserve the mass of active particles within the drop (as shown in Methods). This leads to interesting droplet dynamics including transitions from stationary to motile and back to stationary as a function of activity (see Results: Active Polar Fluid). Secondly, we show that this method can also consistently model an active polar fluid droplet and compare the results of these to previous studies in the literature as well as characterising previously unpredicted states. Furthermore, we have made improvements to previous fluid immersed boundary models by ensuring conservation of droplet area (to be consistent with incompressibility) and to reduce numerical instabilities in the boundary due to the finite sized mesh (as shown in Methods). These are necessary improvements in order to model these active systems where flows are generated locally.}

{ The structure of the paper is as follows.} We first introduce the analytical equations describing the system in the following section, and then the numerical implementation of these is presented in the Methods section. Following this we present tests of the simulations and compare with other simulation data. We then present some interesting results from these simulations for the two cases of active droplets discussed. These include spontaneous symmetry breaking leading to steady swimming states, droplet deformation, and oscillating states.

In the Discussion section we describe potential uses for these simulations and extensions to these that could be performed in the future to investigate even more biologically relevant systems. In particular we discuss various ways that confinement can be added to this model. {Furthermore, we outline how to extend these simulations to 3 dimensions and how we expect the results presented here to generalise to the 3D case.} Finally, we also discuss potential experimental realisation of these simple systems \emph{in vitro} and { how these simulations can be used to simulate multicomponent active droplets or droplets with active and passive components.}

\section*{Model}
%\label{sec:gequations}

\subsection*{Fluid Droplet Model}
%\label{sec:IBanalytics}

We construct numerical simulations of active fluid droplets using an Immersed Boundary (IB) method for modelling fluid-fluid interactions. The IB method was originally formulated for modelling fluid-structure interactions in biological problems, where the structure can be represented by elastic fibres \cite{Peskin1982, Peskin2002}. In our simulations, we follow the work of \cite{Lai2008} which adapts this to fluid-fluid interactions in a 2D domain. We begin by considering an infinitely thin, closed interface $C$ represented by the position vector $\vect{X}=(X(s,t),Y(s,t))$ where $s$ is a Lagrangian coordinate along the curve $C$ and $t$ is time. The force density on this interface due to the surface tension is given in 2D by:

\begin{align}
\label{Fb2D} \vect{F}_b(s,t) = \nabla_s \sqb{\gamma(s,t)\vect{\tau}(s,t)}
\end{align}
where $\gamma$ is the surface tension, $\tau$ is the tangent vector to the surface and $\nabla_s=\partial/\partial s$. This acts as an external force on the fluids, which can be translated to a 2D force density via the 2D Dirac Delta function as follows

\begin{align}
\label{fext} \vect{f}_b(\vect{x},t) = \int_{C} \vect{F}_b \delta(\vect{x} - \vect{X})  {\rm d}s \, 
\end{align}
where $\vect{x}$ is the position vector in the medium. The 2D Dirac Delta is defined by the identity

\begin{align}
\label{delta} \int_{\Omega}g(\vect{x})\delta(\vect{x} -  \vect{X}){\rm d}^2\vect{x} = g(\vect{X}) \, ,
\end{align}
where $\Omega = \Omega_0 + \Omega_1$ is the total fluid domain, consisting of fluids labelled 0 and 1. In the case of incompressible Stokes' flow, which we assume here, force balance in the fluid dictates:

\begin{align}
\label{fbfluid} \eta \nabla^2 \vect{v} &= \nabla P - \vect{f}_b \\
\label{incomp} {\rm where} \quad \divv \vect{v} &= 0  \, ,
\end{align}
where $\eta$ is the fluid viscosity, $\vect{v}$ is the flow velocity and $P$ the pressure. { At the length and velocity scales of the cell cytoskeleton, inertia is negligible. An estimate of the Reynolds number using the parameter tables in the appendix and the density of water $\rho_w=1000$kg m$^-3$ gives $Re\approx \rho_w\act R_0^2/\eta_0^2\approx10^{-11}$ (assuming the maximum velocity scale is $v\approx\act R_0/\eta_0$).} 

Finally, if we assume that the fluids are completely immiscible, then the interface velocity will be the same as the velocity in the medium at that point, such that:

\begin{align}
\label{vint} \pdiff{\vect{X}}{t} = \vect{V} \equiv \int_\Omega \vect{v}(\vect{x},t)\delta(\vect{x} - \vect{X}){\rm d} ^2 \vect{x} \, .
\end{align}
These simple equations model the evolution of a Newtonian fluid droplet embedded in another Newtonian fluid, with a varying surface tension $\gamma(s,t)$ at the interface. In \cite{Lai2008}  surface tension decreases with surfactant concentration \cite{Stebe1996, James2004, Lee2006a}, however here we consider the case where it depends on a concentration of active matter on the interface generating contractile stresses. Thus the model is similar to models of anti-surfactants on an interface that locally increase surface tension \cite{Masoud2013, Conn2016}. The difference in an active system is that { energy is consumed locally by the particles and converted to work, and it is assumed that the chemical fuel concentration can be treated as a constant everywhere}. 

\subsection*{Active Isotropic Interface}
\label{sec:ABanalytics}

We consider an active liquid crystal dispersed on the interface such that the fluid particles are ordered along the tangent line of the interface. This is the 2D limit of the case of a 3D fluid droplet with an isotropically ordered liquid crystal confined to the plane of the interface (the 3D case is discussed { analytically} in \cite{Whitfield2016a}). As in the model of the actin cortex in \cite{Joanny2013} we consider that this active fluid is contractile enough to generate a negative pressure in the (1D) boundary layer for small concentrations, and that a passive pressure will become stronger at larger concentration. This is in qualitative agreement with experiments on reconstituted crosslinked actomyosin \emph{in vitro} which, within a range of myosin and crosslinker density, initially contracts and increases in density until a steady state size is reached \cite{Bendix2008}. In this case, we consider the lowest order contribution to be linear in $c$ as we are using a one-component description. These contractile forces, when confined along the tangent line of the droplet interface, give rise to a change in surface tension equivalent to the stress in the active fluid:

\begin{align}
\label{ST} \gamma(s,t) = \gamma_0 - \actc c(s,t) - \frac{B}{2} c(s,t)^2 \, ,
\end{align}
where $\gamma_0$ is the bare surface tension of the fluid drop. The activity $\actc$ is a phenomenological parameter quantifying the strength of the active contractile $\actc<0$ or extensile $\actc>0$ forces with passive repulsion force proportional to $B \ge 0$. It will be useful to consider results with respect the effective activity $\actt = \actc + Bc_0$ where $c_0$ is the stationary steady state concentration. Note also that there is potentially a passive pressure term linear in $c(s,t)$ absorbed into $\actc$, but as all of these terms are phenomenological this does not change the model. In general the order of the active term in powers of $c$ requires details of the microscopic interaction responsible, but for the scope of this paper we consider this linear case only. As in \cite{Lai2008} conservation of mass on the moving interface $C(t)$ gives:

\begin{align}
\label{advection} \diff{}{t} \int_{C(t)} c(s,t) {\rm d}l(s,t) = \int_{C(0)} \pdiff{}{t} \rndb{c(s,t) \norm{\pdiff{\vect{X}(s,t)}{s}}} {\rm d}s = 0
\end{align}
where $\norm{\partial \vect{X}(s,t)/\partial s}$ is the stretching factor, relating the length of a boundary element ${\rm d}l$ to its initial length ${\rm d}s$ such that ${\rm d}l(s,t) = \norm{\partial \vect{X}(s,t)/\partial s}{\rm d} s$. Eq \eqref{advection} implies that:

\begin{align}
\label{advection2} \pdiff{c(s,t)}{t}  + \frac{c(s,t)}{\norm{\partial \vect{X}(s,t)/ \partial s}}\pdiff{}{t}\norm{\pdiff{\vect{X}(s,t)}{s}} = 0 \,.
\end{align}
The second term in Eq \eqref{advection2} is the Lagrangian advection term. Diffusion of the mass on the closed interface is incorporated with the addition of the term $\partial c/\partial t = D \partial^2 c/\partial l^2$ where $D$ is the diffusion coefficient on the interface. Including this term, the evolution of the concentration $c$ becomes:

\begin{align}
\label{ad_diff} \pdiff{}{t} \rndb{c(s,t) \norm{\pdiff{\vect{X}(s,t)}{s}}} = D \pdiff{}{s}\rndb{\frac{\partial c(s,t)/\partial s}{\norm{\partial \vect{X}(s,t)/\partial s}}} + q(s,t)\,.
\end{align}
The final term $q(s,t)$ captures any binding to or unbinding from the boundary, which we define in Eq \eqref{Q}.

We explicitly consider the coupling of the interface concentration to a bulk concentration $\rho$ in the internal fluid. We only consider this concentration as active when bound to the interface, because an isotropic stress generates no flow in an incompressible medium. In the case of the cytoskeleton active fluid, our system could model a two component system of motor proteins and actin filaments, where the isotropic filament network is concentrated on the interface and remains fixed while a population of motor proteins in the droplet can bind and unbind from this network. Then the equations describing an internal bulk concentration can be incorporated into this model by defining a level-set function $H$ such that:

\begin{align}
\label{Hdefa} \nabla H(\vect{x},t) \equiv -\int_{C(t)} \hat{\vect{n}}(s,t) \delta\sqb{\vect{x} - \vect{X}(s,t)} {\rm d} l = 
\begin{cases}
1 \qquad {\rm if} \quad \vect{x} \in \Omega_0\\
0 \qquad {\rm if} \quad \vect{x} \notin \Omega_0
\end{cases} \, .
\end{align}
Note that length element ${\rm d} l = \norm{\partial \vect{X}/\partial s}{\rm d}s$ changes as the boundary deforms. Then, following \cite{Chen2014} we define the concentration field of the unbound active particles as $H\rho$ and hence the advection-diffusion equation describing this quantity is:

\begin{align}
\label{dcfdt} \pdiff{\sqb{H(\vect{x},t)\rho(\vect{x},t)}}{t} + \divv{\rndb{\vect{v}(\vect{x},t)H(\vect{x},t)\rho(\vect{x},t)}} = D_f\divv{\rndb{H(\vect{x},t)\nabla \rho(\vect{x},t)}} - Q(\vect{x},t)
\end{align}
where:

\begin{align}
\label{Q} Q(\vect{x},t) = \int_C q(s,t)\delta(\vect{x}-\vect{X})\norm{\pdiff{\vect{X}}{s}}{\rm d}s \equiv \int_C \sqb{k_{\rm on} \rho_b(s,t) - k_{\rm off}c(s,t)}\delta(\vect{x}-\vect{X})\norm{\pdiff{\vect{X}}{s}}{\rm d}s \, . 
\end{align}
The unbound concentration of active material at the boundary $\rho_b$ therefore is defined as:

\begin{align}
\label{Cf} \rho_b(s,t) = \int_{\Omega} H(\vect{x},t)\rho(\vect{x},t)\delta(\vect{x}-\vect{X}){\rm d}^2\vect{x}
\end{align}
We assume for this model that the binding rates obey a linear scaling with local concentration. However it is likely that this is not a realistic assumption given the large gradients in concentration field that the simulations predict. Investigating and incorporating more realistic binding behaviour goes beyond the scope of this work and would demand a model { informed by the microscopic details of the system in question}.\\

Eq \eqref{dcfdt} ensures that the total mass of active material is conserved such that:

\begin{align}
\label{ctotcons} \int_{\Omega} \pdiff{\sqb{H(\vect{x},t)\rho(\vect{x},t)}}{t}{\rm d}^2\vect{x} + \int_C \pdiff{c}{t} \norm{\pdiff{\vect{X}}{s}}{\rm d}s  = 0 \,.
\end{align}
Thus these equations describe advection-diffusion inside the droplet with containing boundary conditions.

\subsection*{Active polar fluid droplet}

\begin{figure}[h!]
	\centering
	\includegraphics[width=0.7\textwidth]{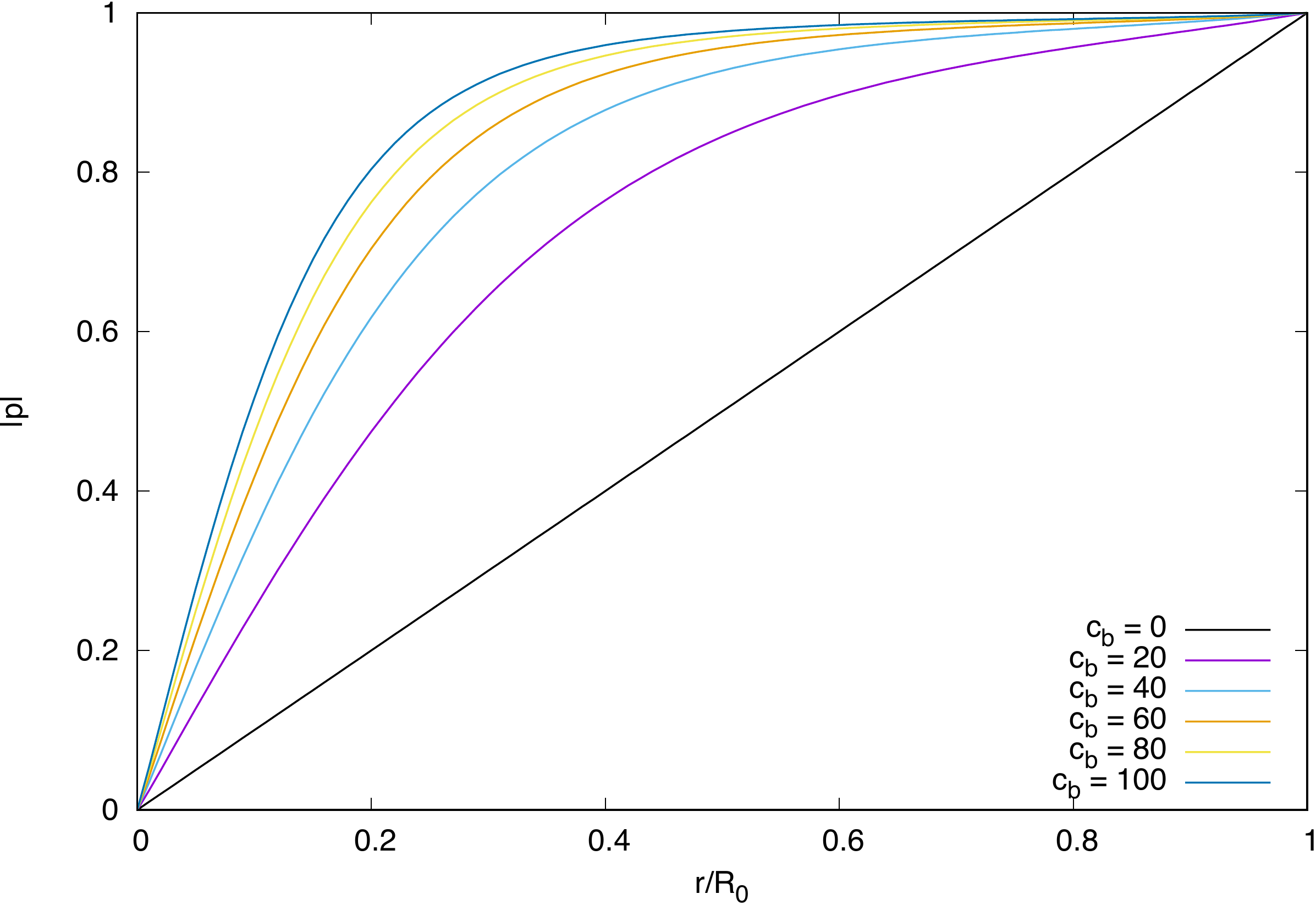}
	\caption{Numerical solutions to Eq \eqref{pmageq} for various values of $c_b$. As $c_b$ increases the distribution tends towards $\norm{p}=1$ everywhere.}
	\label{fig:rdep}
\end{figure}

We independently incorporate an active fluid in the bulk of the drop. As { mentioned in the previous section}, an isotropic active fluid cannot generate flow { if it is} incompressible, so we consider the case of an active liquid crystal with strong polar ordering. Thus, we define a unit polarisation vector $\vect{p}$ which describes the direction of local polar ordering. As shown in \cite{Kruse2004} we can model the passive dynamics of this system by the Frank free energy functional with appropriate boundary terms. In this context we can write this as:

\begin{align}
\label{FreeEH} F_{H} =  \int_{\Omega} f_H {\rm d}^2\vect{x} = \frac{1}{2}\int_{\Omega} \sqb{K(\nabla \vect{p})^2 + \frac{K}{2}c_b\vect{p}^2\rndb{\vect{p}^2-2H} + W \rndb{\norm{\nabla H} + \vect{p}\cdot\nabla H}^2}{\rm d}^2\vect{x} \,.
\end{align} 
The first term in Eq \eqref{FreeEH} is the one constant approximation of the Frank free energy density. The second term is used to replace the Lagrange multiplier which ensures $\norm{\vect{p}}=1$ as used in \cite{Giomi2014}. This term is minimised when $\vect{p}$ is a unit vector and so for large $c_b$ this is realised everywhere inside the droplet except near to any topological defects. This can be seen by writing the polarisation as $\vect{p} = \norm{p}(r)\sqb{\cos(\beta(r,\theta))\hat{\vect{r}} + \sin(\beta(r,\theta))\hat{\vect{\theta}}}$ so that in the large anchoring limit the free energy is minimised for $\beta=0$ everywhere, and $\norm{p}$ must satisfy the Ginzburg-Landau equation \cite{Ginzburg1958}:

\begin{align}
\label{pmageq} \norm{p}''(r) + \frac{1}{r}\norm{p}'(r) - \frac{1}{r^2}\norm{p}(r) = c_b \norm{p}(r)\rndb{\norm{p}^2(r)-1} \, .
\end{align}
This ordinary differential equation (ODE) is non-linear due to the $\norm{p}^3$ term and hence we are only able to solve it numerically. Assuming that $\norm{p}=1$ at the droplet boundary (strong anchoring) and $\norm{p}=0$ at the origin we find the solutions plotted in Fig \ref{fig:rdep}. In the limit $c_b \rightarrow \infty$ this term fulfils the role of the Lagrange multiplier $h_\parallel^0$ and strictly enforces $\norm{p}=1$ on $\Omega_0$ and $\norm{p}=0$ on $\Omega_1$. In the simulations, we have to use finite $c_b$, hence there is some finite gradient of polarisation across the interface and around the defect, which sets the effective core size of the defect in the system.

The final term in Eq \eqref{FreeEH} is minimised for filament polarisation $\vect{p}$ aligned with the outward normal direction near the interface. We consider this polar boundary condition as actin networks are found to order in such a way near the cell membrane in experiments \cite{Small1978}. This term is the source of polar ordering in the simulations as all the other terms are symmetric for $\vect{p} \rightarrow -\vect{p}$.\\

This model now looks similar to many phase field models of (active and passive) liquid crystal emulsions in the literature \cite{Yue2004a, Ziebert2011, Tjhung2012, Giomi2014}. The key difference in the immersed boundary formulation is that the interfacial width does not depend on the surface tension and is a numerical approximation to a sharp interface, and the level-set function $H$ is enslaved to the position of this interface. This means we can usefully change the properties of the boundary without requiring a change to any other part of the calculation. 

Thus, if we assume that activity is confined to the drop we can write the active stress as \cite{Kruse2004}

{
\begin{align}
\label{Hact} \sigma^{\rm act}_{ij} = -\act H \rndb{p_i p_j - \frac{1}{2}\delta_{ij}} \,.
\end{align}
The passive distortion stress in the liquid crystal is defined in terms of the molecular field $\vect{h}$ \cite{Gennes1993}}

\begin{align}
\label{diststress} \sigma_{ij}^{\mathrm{dist}} &= \frac{\nu}{2} \rndb{p_i h_j + p_j h_i} + \frac{1}{2}\rndb{p_i h_j - p_j h_i} + \sigma_{ij}^{\mathrm{e}} \, ,
\end{align}
The molecular field $\vect{h} = -\delta { F_H}/\delta\vect{p}$ acts to minimise the free energy functional $F_H$ with respect to $\vect{p}$, and $\nu$ is the coupling constant between the flow gradient and polarisation. The final term is the Ericksen stress, a generalisation of the hydrostatic pressure for complex fluids, which in this case is given by \cite{Furthauer2012}:

\begin{align}
\label{estress} { \sigma_{ij}^{\mathrm{e}} = { f_H}\delta_{ij} - \pdiff{f_H}{\rndb{\partial_j p_k}} \partial_i p_k} \, .
\end{align}
In addition, there is a stress term due to the polarisation across the interface given by (as used in \cite{Tjhung2012}):

{
\begin{align}
\label{sint} \sigma^{\rm int}_{ij} = -\frac{\partial f_H}{\partial H}H\delta_{ij} - \pdiff{f_{H}}{(\partial_i H)}(\partial_j H) \, .
\end{align}}
The stress then is incorporated into the force balance so that Eq \eqref{fbfluid} becomes:

\begin{align}
\label{fbfluidact} \eta \nabla^2 \vect{v} = \nabla P - \divv\rndb{\tens{\sigma}^{\rm act} + \tens{\sigma}^{\rm dist} + \tens{\sigma}^{\rm int}} - \vect{f}_b \, . 
\end{align}
Finally the evolution of the polarisation field is governed everywhere by \cite{Kruse2004}:

\begin{align}
\label{dpdta}  \pdiff{\vect{p}}{t} = -\rndb{\vect{v}\cdot\nabla}\vect{p} - \tens{\omega}\cdot\vect{p} - \nu\tens{u}\cdot\vect{p} + \frac{\vect{h}}{\Gamma}   \, ,
\end{align}
where the vorticity tensor $\omega_{ij} = (\partial_i \vect{v}_j - \partial_j \vect{v}_i)/2$ and strain rate $u_{ij} = (\partial_i \vect{v}_j + \partial_j \vect{v}_i)/2$ act to reorient the polar particles. 

\section*{Results and Analysis}
\label{sec:resan}
 	
In this section we look at some preliminary tests performed to check the accuracy of the simulations and results that are pertinent to the research questions posed in the introduction. Namely, we quantify symmetry breaking transitions and steady state behaviour including droplet motility and deformation which resemble aspects of cytoskeletal dynamics driven by active stresses. We also look at the role of certain parameters on the phase transitions we observe and relate this to similar systems discussed in the literature.

We normalise the length and time scales of the simulations with respect to the initial droplet radius $R_0$ and the viscosity $\eta$ and following some checks of accuracy choose step sizes of $h=0.075$, $\Delta s \approx h/2$ and $\Delta t = 5\times 10^{-4}$. We choose a grid length of $L_x = L_y = 9$, which gives $M_x = M_y = 120$, which has prime factors of $2$, $3$, and $5$, resulting in efficiency savings in the FFT algorithm. A full list of the parameter value ranges used in these simulations can be found in the methods section. We relate these values to SI units by setting the length, time and energy scales to cytoskeleton relevant values such that $R_0 = 10\,\mu {\rm m}$ and $\eta = 10 \, {\rm kPa\,s}$ \cite{Wottawah2005}, and $\act = -10\ldots10 {\rm k Pa}$ \cite{Kruse2006}.

\subsection*{Fluid Droplet Deformation}
\label{sec:shapetest}

We first test the fluid simulation code alone without any active terms to check that it agrees with the expected behaviour of a simple fluid drop in 2D. We consider a droplet with an initially deformed shape given by:

\begin{align}
\label{testshape} \vect{X}^0_k = \sqb{1 + \frac{1}{2}\cos(2\theta^0_{k+1/2})}\sqb{\cos(\theta^0_{k+1/2}),\sin(\theta^0_{k+1/2})}
\end{align}
where $\theta^0_{k+1/2} = ({k+1/2})\Delta s$ (see Fig \ref{fig:testshape}). We can then measure the evolution of this shape at time $n$ using the Fourier coefficients:

\begin{figure}[h!]
	\centering
	\includegraphics[width=0.8\textwidth]{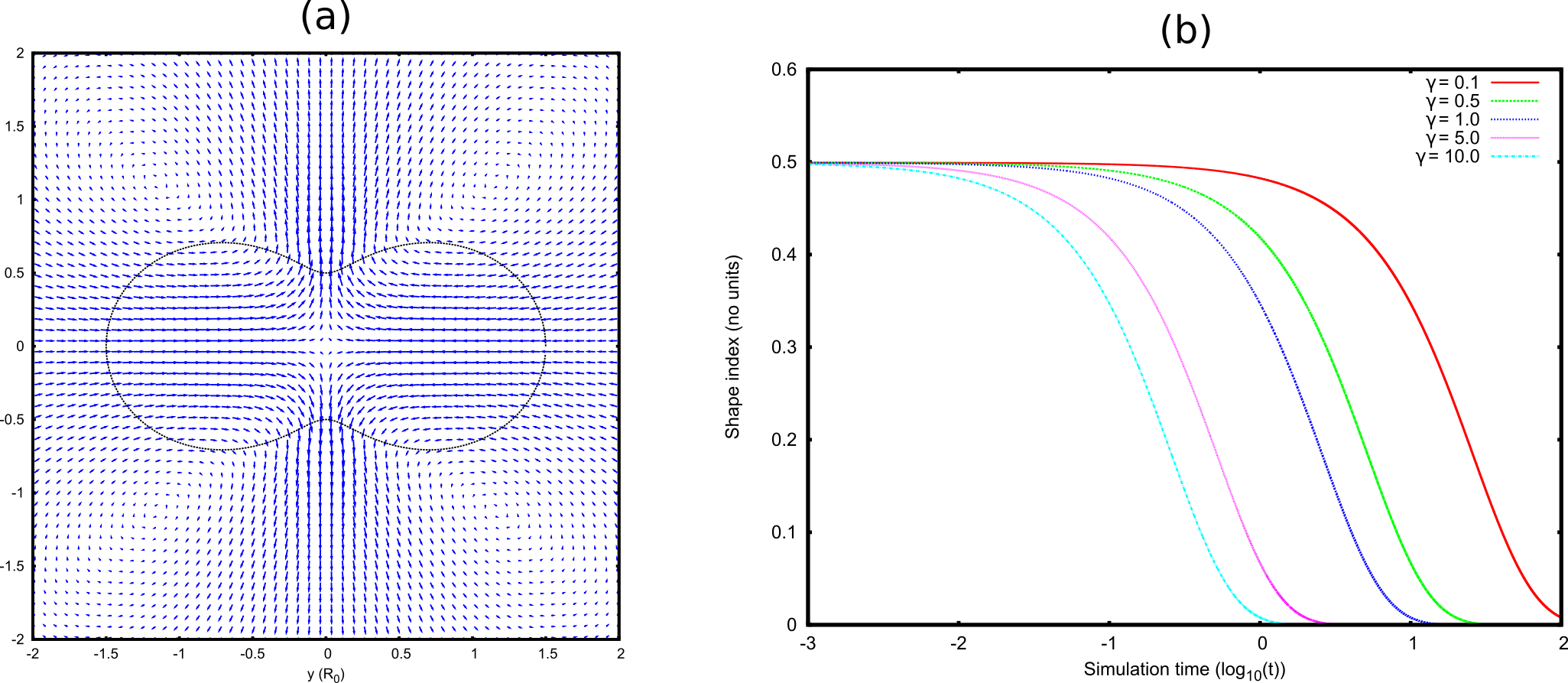}
	\centering
	\caption{\textbf{(a)} Black line shows the boundary conformation at time $t=0$ given by Eq \eqref{testshape}. Blue arrows show the initial flow field, scaled for visibility. \textbf{(b)} Evolution of the shape factor $a^n_2$ (see Eq \eqref{fcoeff1}) with respect to the simulation time $t$ (log scale) for values of the surface tension $\gamma$ (in simulation units) spanning two orders of magnitude.}
	\label{fig:testshape}
\end{figure}

\begin{align}
\label{fcoeff1} a^n_l = \sum_{k=0}^{N-1} \frac{1}{2}\rndb{\norm{\tilde{\vect{X}}^n_k} + \norm{\tilde{\vect{X}}^n_{k+1}}} { \cos\sqb{\frac{l}{2}(\theta_k^n + \theta_{k+1}^n)}} \rndb{\theta_{k+1}^n - \theta_{k}^n}\\
\label{fcoeff2} b^n_l = \sum_{k=0}^{N-1} \frac{1}{2}\rndb{\norm{\tilde{\vect{X}}^n_k} + \norm{\tilde{\vect{X}}^n_{k+1}}} { \sin\sqb{\frac{l}{2}(\theta_k^n + \theta_{k+1}^n)}} \rndb{\theta_{k+1}^n - \theta_{k}^n}
\end{align}
{ where $\tilde{\vect{X}}$ is the point position in the coordinate system with the droplet centre of mass at the origin and $\theta_k^n = \tan^{-1}(\tilde{Y}_k^n/\tilde{X}_k^n)$}.  Generally, we compute the first few orders of Fourier coefficients $l=2,3,4$, but in this particular test we just present $a^n_2$ as the shape index as this is initially set to $1/2$ by Eq \eqref{testshape}. The evolution of this shape index is presented in Fig \ref{fig:testshape}(b) for several values of surface tension $\gamma$. As one would expect the droplet returns to a spherical shape, and the time it takes to do so is linearly proportional to the surface tension. Reducing the grid spacing from $h=0.075$ to $h=0.025$ makes no discernible difference to the droplet evolution and nor does increasing the grid size $L_{x,y}$. Thus our choice of $L_{x,y} = 9R_0$ appears sufficient to render interactions with the droplet's periodic image negligible.
 
We also note that once the droplet reaches steady state, there is still some flow in the system even though analytically the flow should be zero. This arises due to inaccuracies in the mapping to the Cartesian mesh. We find that these flows are very small and do not affect the droplet shape. However, these flow are on average only an order of magnitude smaller than those produced by the random perturbation that we apply, so they could affect the symmetry breaking direction.

\subsection*{Active Boundary Simulations}

In this section we present the steady states observed in the active isotropic interface model. First, we  consider the case where the active fluid is bound to the interface, setting $k_{\rm off} = 0$ and quantify the steady states observed. Following this we increase this binding rate and observe the differences this coupling can make to the observed steady state.

In all of these cases, in order to enable the droplet to break symmetry we apply a perturbation to the system at time $t=2.5$. We do this by altering the concentration at the interface at the beginning of the time-step by a small amount. In order to conserve mass, we perturb the concentration using a Fourier series:

\begin{align}
\label{pert} c_k^{5000} = c_k^{5000} + c^0\sum_{l=1}^{N/4} \sqb{ a_{kl} \cos(l\theta_k) + b_{kl}\sin(l\theta_k)}
\end{align}
where $\theta_k$ is as defined in { Eqs \eqref{fcoeff1} and \eqref{fcoeff2}} and the coefficients $a_{kl}$ and $b_{kl}$ are assigned randomly generated values between $-10^{-3}$ and $10^{-3}$.

\subsubsection*{Activity and diffusion only}

\begin{figure}[h!]
	\centering
	\includegraphics[width=0.6\textwidth]{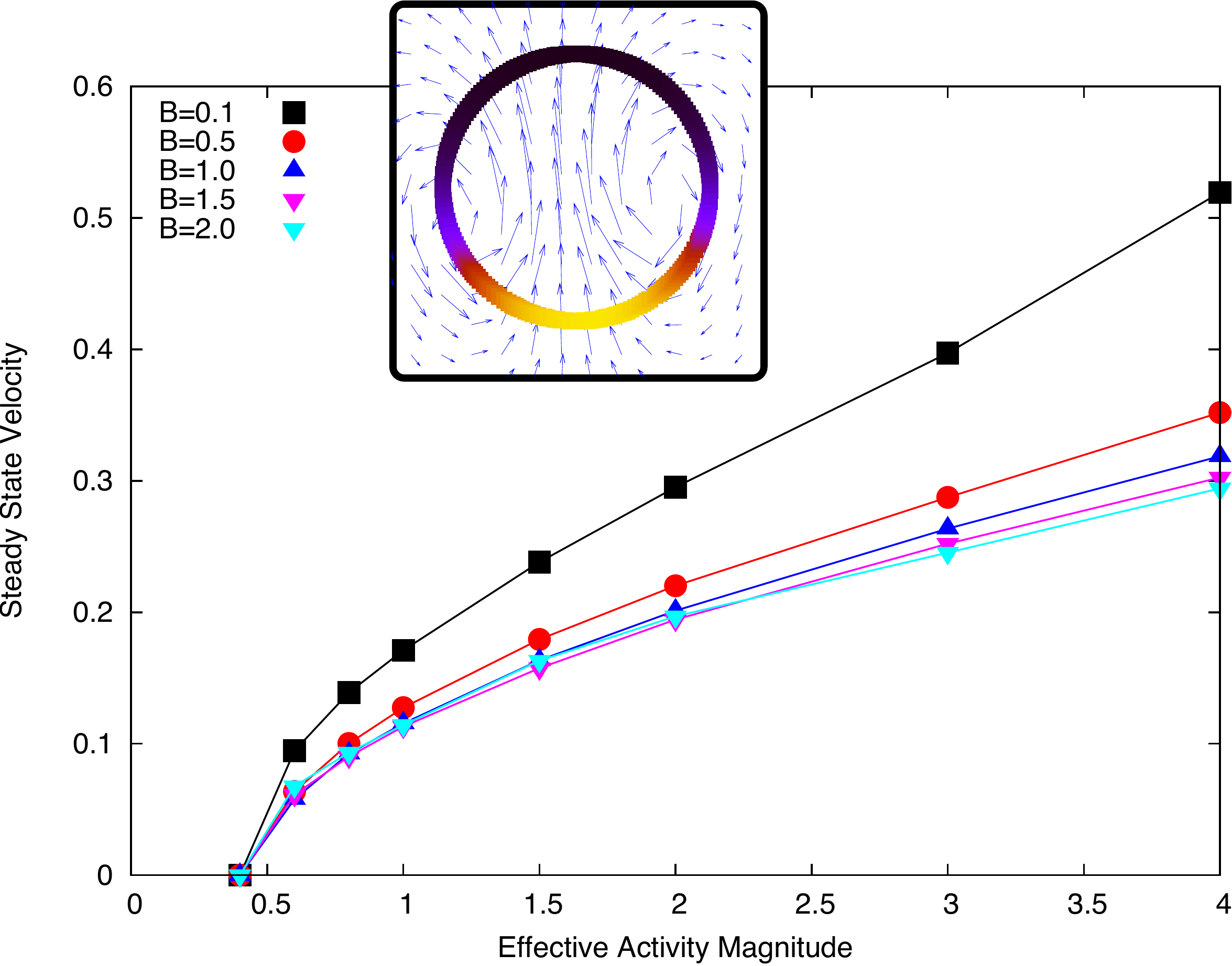}
	\caption{Final droplet velocity plotted against the effective activity $\actt$ for different values of $B$. Inset: Droplet swimming steady state, coloured interface shows concentration (black to yellow: low to high) and vector field shows flow field scaled for visibility. { Simulation parameters: $D=0.1, \gamma_0=\eta=R_0=c_0=1$.}}
	\label{fig:abBspeed}
\end{figure}

We find that for all values of contractile activity tested above a threshold value, a swimming steady state is observed with the droplet remaining circular (see Fig \ref{fig:abBspeed}) as we predict analytically in \cite{Whitfield2016a}. We find it necessary to use a non-zero value of $B$ on the interface in order to ensure numerical stability and physically reasonable results. In Fig \ref{fig:abBspeed} we show how the dependence of steady state droplet swimming speed against effective activity $\actt$ only depends weakly on the parameter $B$. The effect on droplet speed of the parameter $B$ is weak because while increasing $B$ reduces the magnitude of the gradients in $c$ (and hence $\gamma$) along the interface, these gradients are spread over more of the droplet interface and still generate flow towards the concentration peak.

We also see from Fig \ref{fig:abBspeed} that the activity dependence of the speed is typical of active steady state phenomena \cite{Tjhung2012} initially showing a linear dependence on activity near the threshold and a weaker dependence at higher activity values. This suggests an optimum swimming speed efficiency, above which extra swimming speed begins to cost more active energy.

\begin{figure}[h!]
	\centering
	\includegraphics[width=\textwidth]{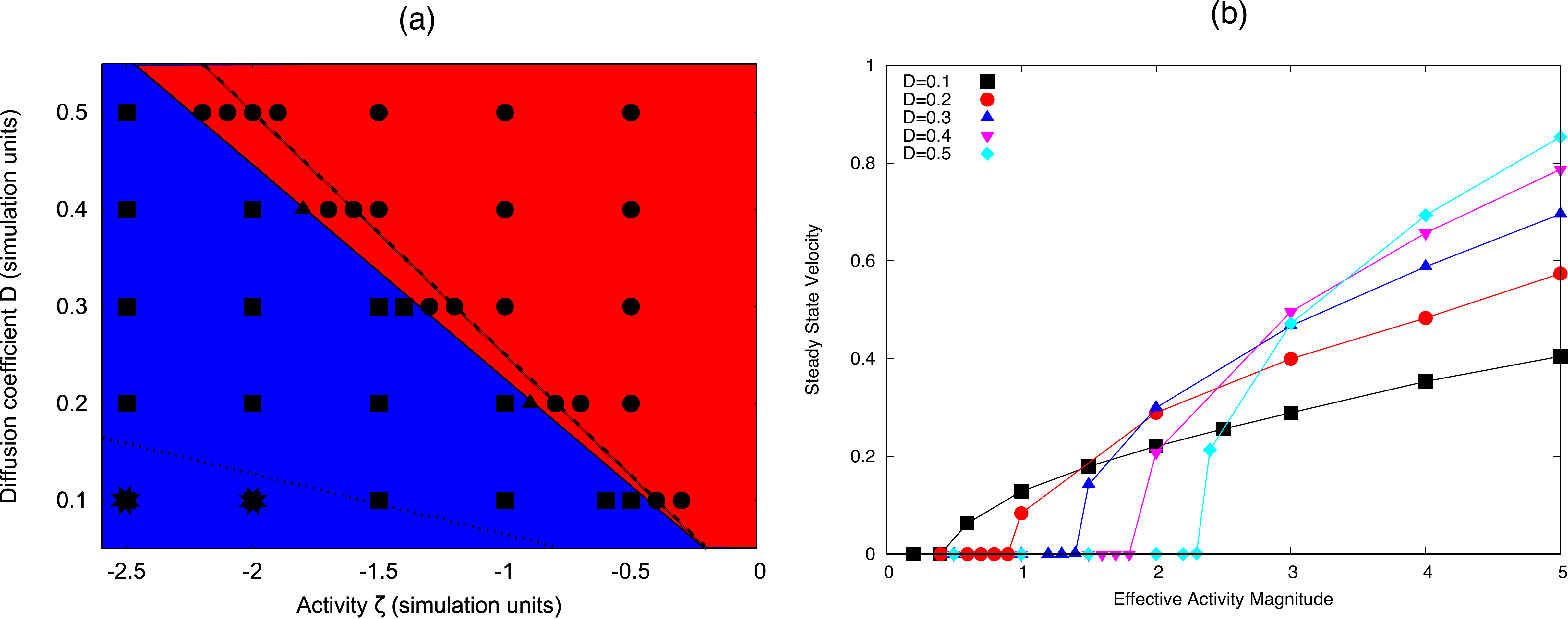}
	\caption{(a) Phase diagram for transition to motile phase with respect to activity and diffusion parameters on the interface. The red region (circles) show simulations where the droplet remains stationary, the blue region (squares) shows simulations where the droplet swims with a single concentration peak. The simulations indicated by stars show where initially the concentration forms two peaks before coalescing into the stable swimming state. (b) Steady state swimming speed versus active for difference values of the diffusion coefficient $D$.  { Simulation parameters: $B=0.5,\gamma_0=\eta=R_0=c_0=1$.}}
	\label{fig:ABphase} 
\end{figure}

The phase diagram in Fig \ref{fig:ABphase}(a) shows how the activity threshold depends on the diffusion of $c$ on the interface. This dependence is approximately linear and is well characterised by the Linear Stability Analysis presented in \cite{Whitfield2016a}. Interestingly, Fig \ref{fig:ABphase}(b) shows that even though diffusion increases the active energy required to break symmetry, faster diffusion rates result in faster swimming droplets at the same activity values. Further, we see that at large enough activity, we observe the formation of two peaks in the concentration profile, which quickly coalesce to form one.

\subsubsection*{Coupling to a bulk concentration}
\label{abdiffbulk}

\begin{figure}[h!]
	\centering
	\includegraphics[width=\textwidth]{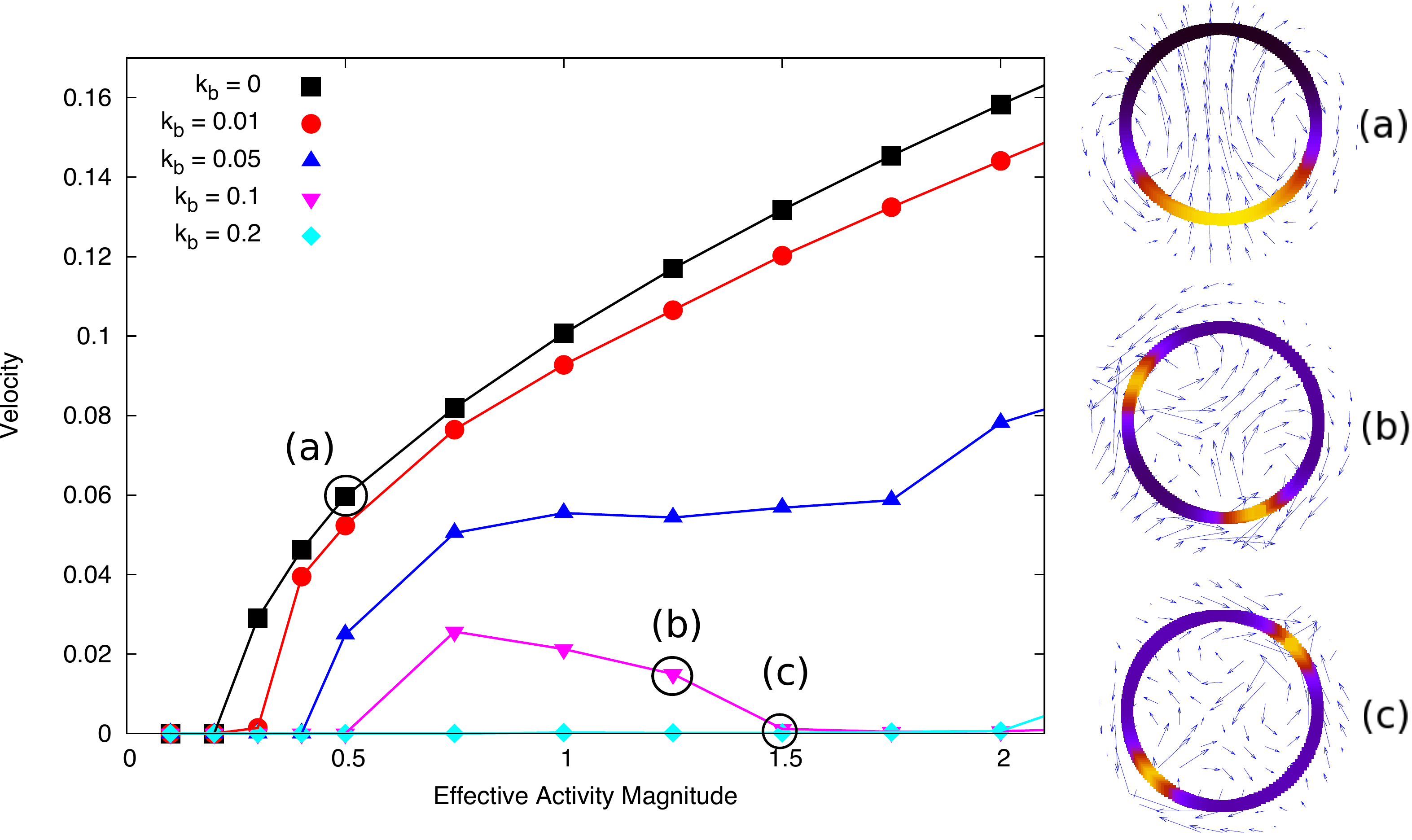}
	\caption{Steady state velocity versus $\norm{\actt}$ over a range of contractile activity values for different binding rates $k_b=k_{\rm off}=k_{\rm on}$. Insets show steady state configurations with concentration on interface shown by colour density (black to yellow) and blue arrows show the velocity. Higher binding rates reduce the droplet speed and enables the formation of two concentration peaks (see insets (b) and (c)). { Simulation parameters: $D_b=0.5$, $D=0.05$, $B=0.5$, $\gamma_0=\eta=R_0=c_0=1$.}}
	\label{fig:bulkvel}
\end{figure}

Setting the binding rates $k_{\rm on,off}$ to be non-zero can completely alter the dynamics of the system. In the case of slow bulk diffusion or binding rate compared to the active flow timescale, we see that this coupling simply dampens the instability discussed in the previous section and reduces the final droplet speed. This is because the active concentration is recycled by the internal fluid, unbinding more at peak concentrations and diffusing slowly through the drop (see Fig \ref{fig:bulkvel}). We fix the bulk diffusion $D_b$ and increase the binding rate $k_b=k_{\rm on}=k_{\rm off}$ and observe that the feedback from the internal concentration changes the observed behaviour. The flow field in the swimming drop reference is rearward near the interface and forward through the middle. This means that the active particles unbinding at the droplet rear are advected through the middle of the drop and this feedback can lead to the nucleation of a second peak in the concentration. In the simulations this results in a continuous transition with respect to $\actt$ from a steady motile state to `wandering' motile states (characterised by the periodic nucleation and coalescence of a second concentration peak, snapshot shown in Fig \ref{fig:bulkvel}(b)) and stationary two-peak solutions (Fig \ref{fig:bulkvel}(c)). Movies of these simulations can be found in the Supplementary Information. 

We see that the wandering state is really a transition phase from steady motility to a stable two-peak configuration. This result is very interesting as it suggests a simple feedback mechanism such as advection-diffusion of material through the droplet is sufficient to stabilise the two-peak solution. Physical intuition suggests that two peaks of this nature should always be unstable due to the hydrodynamic interaction between them. Thus these stable solutions highlight a simple physical mechanism that is generic to such droplet systems. However, as outlined in \cite{Whitfield2016a}, this transition is not simply a function of the effective Pecl\'{e}t number in the drop, since faster diffusion in the drop decreases the activity required to spontaneously form multiple peaks in concentration on the interface. However, it is clear that the binding rate and advection terms need to be large to stabilise the two-peak solution. 

\subsection*{Active Polar Fluid}
\label{sec:apfresults}

\subsubsection*{Comparison with published results}

\begin{figure}[h!]
	\centering
	\includegraphics[width=0.8\textwidth]{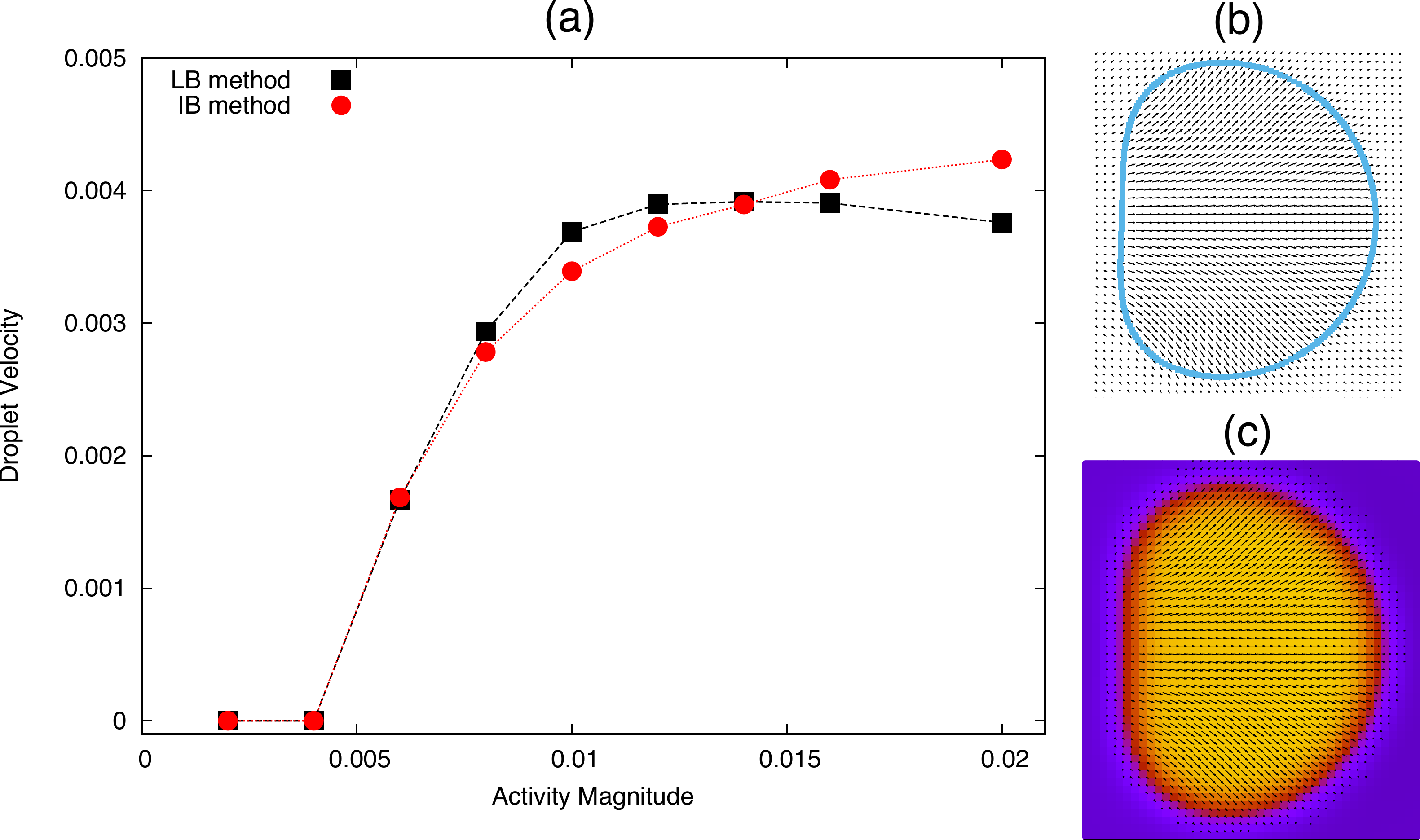}
	\caption{\textbf{(a)} Steady state droplet velocity for an active polar fluid drop simulated using the Immersed Boundary method (red triangles) and using hybrid lattice Boltzmann simulations used in \cite{Tjhung2012}. All parameters used are listed in the text. \textbf{(b)} Snapshot of the system steady state using the IB method, where the blue line is the Lagrangian droplet interface and the black arrows show the polarisation. \textbf{(c)} Snapshot of the system steady state using hybrid LB simulations, the colour gradient shows the value of the phase field $\phi$ {( purple outside, yellow inside)} and the black arrows show the polarisation field.}
	\label{fig:lbcomp}
\end{figure}

We can test that the simulations of a droplet of active polar fluid are working by comparing to the hybrid lattice Boltzmann (LB) simulations published in \cite{Tjhung2012}. By adjusting our simulation parameters to be the same as in the hybrid LB simulations of an active polar fluid we can compare the steady states directly. As in reference \cite{Tjhung2012} we set the following parameters as: $K=0.04$, $c_b = 2.5$, $R_0 = 17$, $\Delta t =1$, $h=1$, $W_1=0$, $\eta=5/3$, $\Gamma = 1$ and $\nu = -1.1$. We initialise both sets of simulations with a small polarisation in the positive $x$ direction inside the droplet, and perturb the systems by applying a small normal anchoring $W_1 = 0.0001$ for the first 5000 time-steps. We set the surface tension to $\gamma = 0.15$ which is approximately the effective surface tension in the phase field model in \cite{Tjhung2012}. We also note that, the active stress in the hybrid LB simulations is proportional to $\act \phi$ where $\phi=2$ inside the droplet, and $\phi=0$ outside. Therefore, when we compare activity values in Fig \ref{fig:lbcomp}(a), the activity values used in the hybrid LB simulations are multiplied by $2$ for consistency.

We can see from Fig \ref{fig:lbcomp} that we have good agreement between the two sets of simulations, with similar steady state velocity trends and droplet shape over a range of contractile activity values. We see that the symmetry breaking threshold agrees well between the simulations, however the non-linear dependence of droplet speed on activity is slightly different. This may in part be due to the estimates of surface tension which have no analytical analogue in the LB simulations, as well as the only approximate incompressibility achieved by LB (which is exact in our model).

\subsubsection*{Symmetry Breaking: Motion and Deformation}

In this section we consider specifically the case with a large polar anchoring term $W_1=50h$ in order to investigate the effect of this coupling between internal polarisation and shape. We observe that with this boundary condition we only see an active motile state for extensile activity $\actt > 0$. This is in contrast to simulations by Giomi \emph{et al.} \cite{Giomi2014} which show that, with normal anchoring an active \emph{nematic} droplet shows swimming states for contractile activity. This suggests that it is the difference between nematic and polar order that is responsible for the difference. The stationary state of the system with polar order consists of an aster with defect $q=+1$ at the droplet centre, whereas in the nematic case it consists of an elongated droplet with $q= \pm1/2$ defects along the long-axis of the droplet. This difference in stationary state therefore changes the geometry such that the stability threshold for motion changes sign. This change appears to be robust for the values of $\nu$ we have tested ($\pm1.1$ corresponding to `rod-like' and one to `disc-like' flow aligning particles \cite{Gennes1993}).

\begin{figure}[h!]
	\centering
	\includegraphics[width=0.7\textwidth]{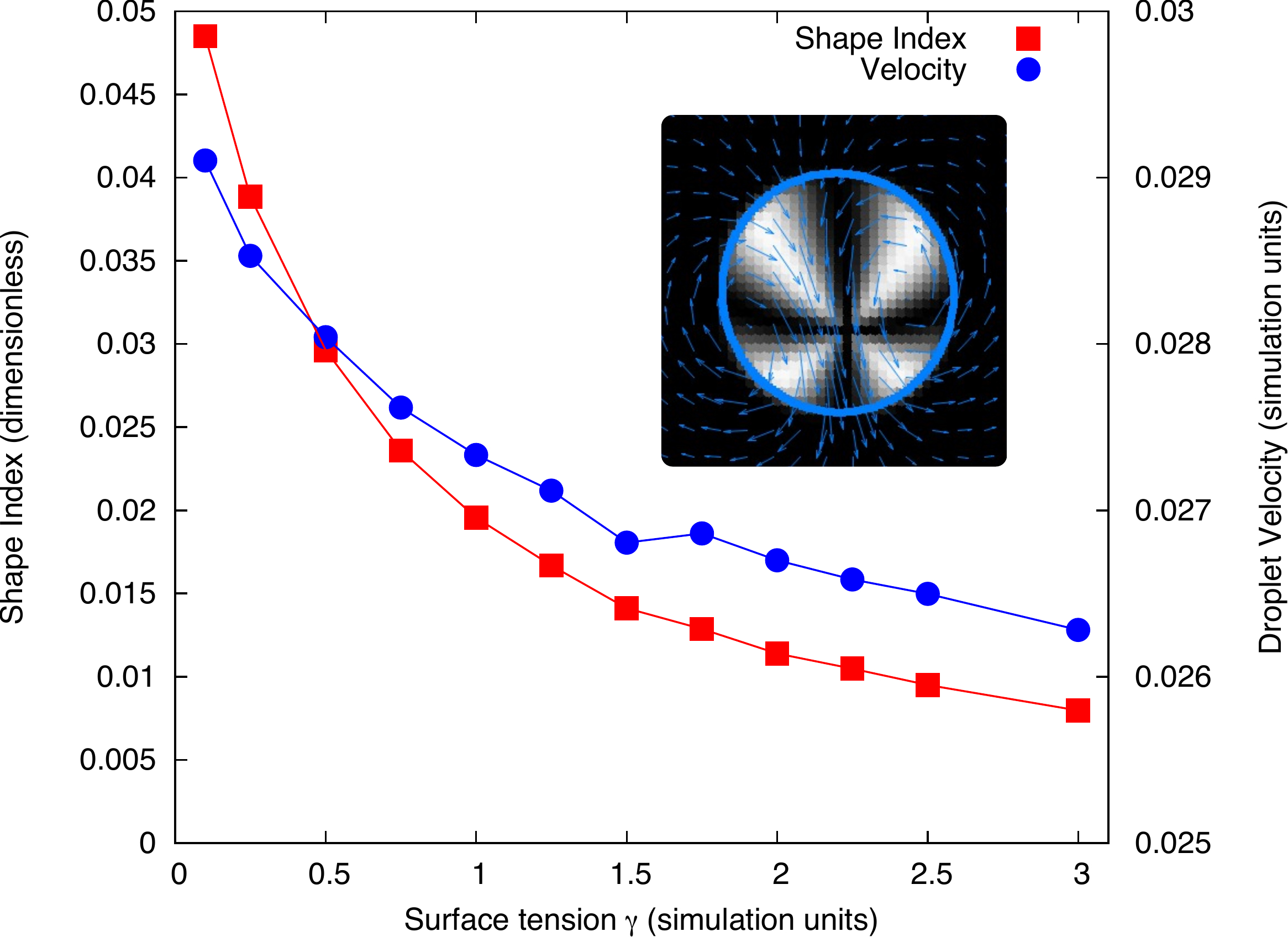}
	\caption{Final droplet speed (right axis) and $k=2$ mode shape parameter (calculated as $\sqrt{(a^T_2)^2 + (b^T_2)^2}$ from Eqs \eqref{fcoeff1} and \eqref{fcoeff2} at time $t=T$, left axis) plotted against droplet surface tension. The simulations use $\act = 1.5$ and $\nu=1.1$. Inset shows the swimming steady state for $\gamma=1$ where the shading is a Schlieren texture representation of the polarisation field for crossed $x$-$y$ polarisers such that dark regions represent areas where the polarisation is parallel or antiparallel to the $x$or $y$ direction (or isotropic as is the case outside the droplet). The blue arrows show the flow field in the fluid (averaged over neighbouring points for clarity), while the blue line traces the droplet interface. Other parameters were set as listed in tables \ref{Fluid params} and \ref{APFparams} with $K=0.1$.}
	\label{fig:shapevel}
\end{figure}

As predicted in \cite{Whitfield2016a} the activity threshold for the onset of motion does not depend on the surface tension, however the droplet steady state speed does. As we can see in Fig \ref{fig:shapevel} the motile steady state is elongated and the shape deformation depends on the surface tension. In Fig \ref{fig:shapevel} we plot the $k=2$ mode shape parameter $\sqrt{(a^T_2)^2 + (b^T_2)^2}$ from Eqs \eqref{fcoeff1} and \eqref{fcoeff2} (essentially measuring droplet elongation) and final droplet velocity against the surface tension. Clearly, due to the strong anchoring at the interface, the polarisation profile is coupled to the boundary shape, so when the surface tension is small the polarisation profile is able to deform the droplet more and reach a faster steady state. As surface tension is increased the deformation reduces and goes towards a circular droplet, and in this limit the droplet velocity begins to plateau at its minimum value. This shows a positive feedback between the shape and the motion in these droplets, but in this case it does not affect the symmetry breaking threshold. { This defect driven motion is similar in appearance to those predicted for crawling active droplets in \cite{Tjhung2015,Khoromskaia2015}, despite the difference in geometry and flow boundary conditions.}

\begin{figure}[h!]
	\centering
	\includegraphics[width=\textwidth]{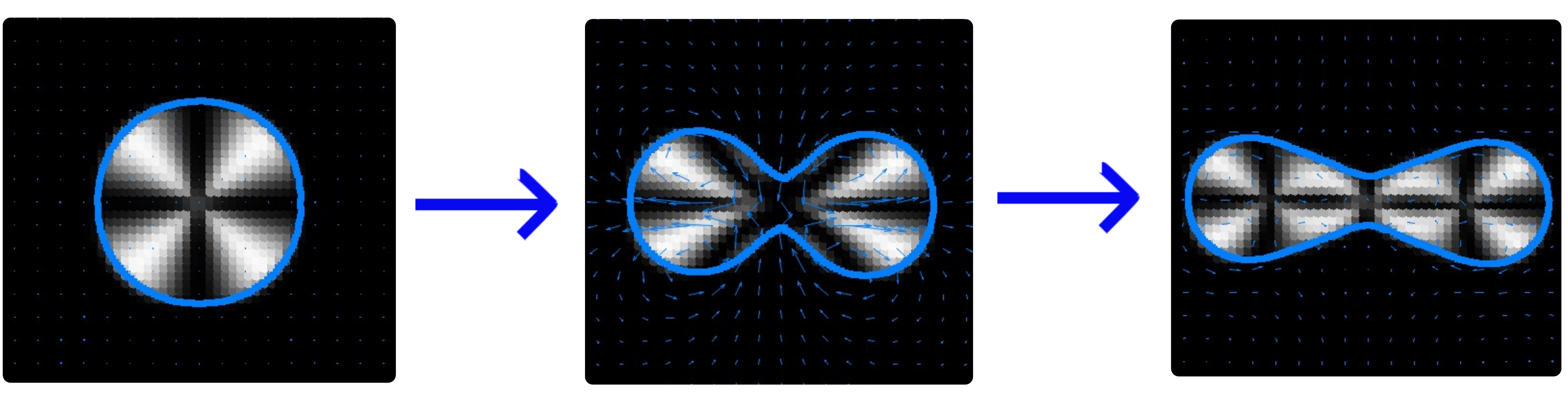}
	\caption{Snapshots of the simulations for contractile activity $\act=-4$ and surface tension $\gamma=0.1$ at 3 different times in the simulation (corresponding to $t=10$, $t=50$ and $t=60$ respectively in simulation units) using the same format as the inset in Fig \ref{fig:shapevel}. All other parameters are the same used to obtain Fig \ref{fig:shapevel}.}
	\label{fig:deformed}
\end{figure}

For contractile activity we see that the lowest order symmetry breaking mode corresponds to a non-motile state where the droplet appears to initiate division. The rate of deformation of the droplet interface is exponential and is modelled well by the linear stability analysis at small times. In some cases we see that this deformation is halted at some steady state `dumbbell' shape, where the defect is elongated into a line defect along the long axis of the droplet and can separate into 3 distinct defects (of charge +1,-1 and +1 respectively) shown in Fig \ref{fig:deformed}. At larger activity values or lower $\gamma$ we see that the opposite sides of the immersed boundary are almost brought into contact. However, these simulations do not yet have the capability of modelling division of the droplet as this is not a part of the IB model, which is a shortcoming in comparison to phase field models. Thus in these cases the simulation method breaks down. In most cases we test the deformation does stop and we observe further symmetry breaking of the internal polarisation field such that the droplet front enlarges and the droplet swims with the 3-defect configuration. However we do not characterise these dynamics here as they are complex and depend strongly on the value and sign of the liquid crystal coupling parameter $\nu$.

In principle one could impose conditions that would allow division of the boundary, such that there were two separate immersed boundaries in the simulation. To do this, one would need to calculate the separation between non-neighbouring boundary points $k_1$ and $k_2$, and once they were within a certain distance (say ${\rm d}s$) separate the boundary into 2 boundaries one from $k_1+1\ldots k_2$ the other from $k_2 \ldots k_1$.

\subsubsection*{Onset of Active Turbulence}

It is well known that active systems can display turbulent like behaviour at zero Reynolds' number when the the equilibrium relaxation time is long compared to the timescale of the active dynamics. In terms of activity, the threshold for this behaviour is proportional to $\eta K / { \Gamma} L^2$ where $K$ is the one Frank elastic constant and $L$ is the system size \cite{Ramaswamy2010, Giomi2012, Marchetti2013}. These turbulent-like states have been observed experimentally in bacterial suspensions and reconstituted cytoskeletal networks \cite{Dombrowski2004a, Sanchez2012} where the active flows can spontaneously create and annihilate polarisation defects in pairs and generate vortices in the flow. We observe the onset of such dynamics in an active polar fluid droplet by reducing the elastic constant in our simulations to $K=10^{-3}$. We find that due to the strong anchoring condition at the interface, and relatively high surface tension the polarisation field remains approximately fixed at the boundary. However, reducing the constant $K$ reduces the equilibrium interaction between these filaments and those near the droplet centre which show dynamics that are only weakly coupled to the interface.

\begin{figure}
	\centering
	\includegraphics[width=\textwidth]{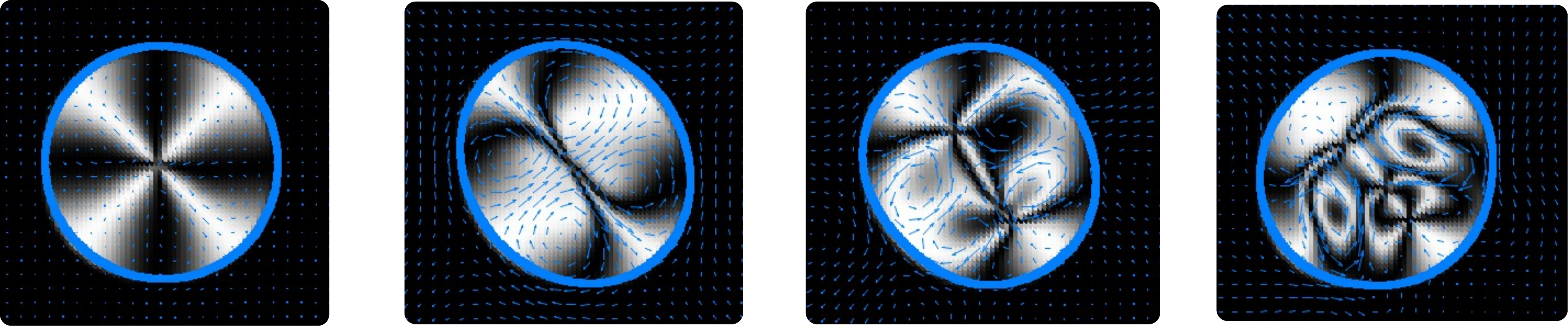}
	\caption{Snapshots of the simulations for contractile activity $\act=-1$, surface tension $\gamma=2$ and $K=10^{-3}$ at 4 different times in the simulation (corresponding to $t=$25, 75, 125 and 250 respectively in simulation units). The shading represents the Schlieren texture as used in Fig \ref{fig:deformed}. The blue arrows show the flow field in the fluid (averaged over neighbouring points for clarity), while the blue line shows the droplet interface. We also use a finer mesh than that used to obtain Fig \ref{fig:deformed}, with $h=0.04$ and $N=150$ in order to resolve the defect dynamics more clearly.}
	\label{fig:turbulence}
\end{figure}

This transition to turbulent motion is easily observed if we consider the case of contractile activity which for low activity values stabilises the defect at the droplet centre. Snapshots of the dynamics are displayed in Fig \ref{fig:turbulence}. We see that due to the low $K$ value the stationary configuration becomes unstable but the flows are not large enough to significantly deform the drop. The result is unsteady dynamics of the polarisation field in the centre of the drop that are mostly decoupled from the interface dynamics. This results in stretching and distortion of the original $+1$ defect into a line and then further pairwise creation of $\pm1$ defects (shown in Fig \ref{fig:turbulence} by the points where four black stripes intersect). As we scale the droplet size up, or decrease $K$, one would expect that these turbulent dynamics would become faster as the effective energy cost of deformation becomes smaller relative to the activity. Note that due to the polar nature of the system modelled here, only defects of integer topological charge can be generated, so these have fundamentally different dynamics to the half integer defects observed in active nematics.

\section*{Discussion}
\label{sec:discussion}

{
\subsection*{Extension to 3 dimensions}
\label{sec:3D}

This method can be extended to a full 3D simulation, but the steps required to do so go beyond the scope of this paper. The equations related to the fixed fluid grid, such as the fluid flow equations \eqref{fbfluidact} and polarisation dynamics equations \eqref{dpdta} remain unchanged in 3D and the same combination of Fast Fourier Transform and Runge-Kutta methods (see Methods) can be utilised to solve for the dynamics of these quantities. However, the fluid-fluid interface is a 2D surface in the generic 3D case, and so this requires a different representation from the 1D Lagrangian mesh used here. A direct extension would be to represent the fluid interface as a 2D triangular mesh. Then the boundary points will evolve in exactly the same way as before. However, rather than removing single points from the interface (as outlined in Methods) one would have to remove individual triangles below a threshold area and replace the mass contained within that triangular area. Similarly, triangles with excessive eccentricity or area should be split along their short axis into two triangles to ensure stability. In general one could use any standard re-meshing algorithm ensuring that the concentration on the interface is numerically conserved and the spatial distribution is unchanged. Such a mesh has been used to model surfactant coated drops in \cite{Yon1998}. 

An alternative method is to only model the Immersed Boundary points implicitly via the Cartesian fluid mesh (known as the Immersed Interface Method \cite{LeVeque1994}). Then one has to explicitly consider the stress jump in the normal direction at the interface in order to impose the boundary force \cite{Xu2014}. This normal force is then imparted on the Cartesian mesh within the interfacial region (defined by the numerical delta function range). Similarly, the boundary region and associated level-set function are advected due to flow normal to the interfacial region. Thus the approach is closely related except that the boundary mesh is implicitly defined rather than an explicit Lagrangian mesh. 

We expect that the dynamics reported in this paper will be representative of what would be seen in 3D. In \cite{Whitfield2016a} a linear stability analysis of these two systems predicts the same qualitative symmetry breaking behaviour despite the flow being quantitatively changed by the geometry. The main qualitative difference one would expect to observe is a deformation of the droplets with an active interface in 3-dimensions, and an investigation of how this deformation couples to concentration dynamics would be an interesting investigation for future work.

\subsection*{Summary and Conclusions}
}

In this paper we have detailed the numerical method that we have used to simulate the evolution of active droplets in 2D. We have used an Immersed Boundary (IB) method to model the fluid boundary as a Lagrangian mesh of points. This allows us to explicitly simulate a concentration of isotropic active contractile particles on the interface that alter the surface tension. Further, we use the immersed boundary to define a level-set function which defines the inside and outside of the drop. Using this we can define an active polar liquid crystal phase in the droplet interior with anchoring at the droplet interface.

{ Our results reveal steady state migrating states in both types of active droplet. In all cases we investigate the steady state droplet velocity, which scales non-linearly with activity, levelling off at larger activity values. This means there is a generic reduction in swimming efficiency as activity is increased in these systems as observed for the flow rate in other active droplet and bulk systems \cite{Tjhung2011}.

The first system we consider }is a contractile isotropic active fluid on the droplet interface and observe that without any binding and unbinding the only observed active dynamic steady state is a swimming droplet with one peak in concentration. We see that for small binding rates the steady state remains unchanged but reduces the droplet velocity. At higher binding rates feedback from the fluid flow stabilises a second concentration peak and the droplet can reach a stationary two-peak configuration for large enough activity. This shows how a simple feedback mechanism can stabilise more complex behaviour, since at low binding rates if two peaks are formed in the concentration they always attract and coalesce to form a single peak at the droplet rear. { These higher order states show that this active interface shares much of its internal dynamics with models of thin layers of active fluids \cite{Bois2011,Hawkins2011}, and so provides a useful insight into how such a system evolves immersed in an external fluid. Thus, it is a useful step in modelling the dynamics of the actin cortex in cells which many active thin film models are based on.}

Secondly, we simulate an active polar fluid droplet with strong anchoring at the interface. We see that if when we take the Frank elastic constant to be significant (or equivalently a small droplet) we see transitions to highly ordered dynamic states, including motion, rotation and symmetric deformation. { Due to the geometry of the droplets and the polar anchoring condition, we find strikingly different behaviour for the contractile and extensile regimes, with extensile activity promoting translational symmetry breaking and contractile promoting symmetric modes of deformation.} Using a smaller Frank constant (equivalent to a larger droplet) we observe the onset of turbulent dynamics, discussed in depth for active nematics in \cite{Giomi2013}. Due to the confinement inside the droplet the dynamics are still constrained and we see that the number of defects in the polarisation field rarely exceeds 3 for the range of activities simulated here. However, the dynamics are unsteady due to the slow relaxation of the molecular field compared to the active flow timescale.

{ The results discussed here are consistent with previous studies of other active fluid droplet systems \cite{Tjhung2012,Giomi2014}. Additionally, given the increase in experimental progress in producing stable active gels and fluids in vitro \cite{Bendix2008}, and confining such systems to droplets and vesicles \cite{Tsai2011,Sanchez2012,Keber2014} it is possible that these simplified models can be tested experimentally. In particular the active fluid interface model may be achievable by coating a fluid droplet or the inside of a vesicle with a stable isotropic crosslinked actin layer. Active contraction of the layer could be introduced by trapping myosin motor complexes inside the droplet/vesicle using existing microfluidic techniques. Such motor complexes may have { different} binding kinetics than the linear case assumed here, however one would expect to observe qualitatively similar behaviour.}

A useful extension of these simulations that is beyond the scope of the work we present here would be to simulate these systems in 3 dimensions. From our analysis in \cite{Whitfield2016a} we would expect much of the observed dynamics to remain qualitatively the same in 3D, with the exception that we would expect deformation of the droplets under surface tension gradients. In particular it would be interesting to simulate the droplet deformation when we predict steady two peak solutions in the concentration of an active isotropic fluid on the droplet interface (see Fig \ref{fig:bulkvel}) which could equally manifest as a ring or two poles of higher concentration in this geometry.

The simulations presented here can also be used to model interactions with external boundaries or elastic solids in a self-consistent way through the IB method. Similarly, one could investigate the dynamics of multiple active droplets simply by adding more Lagrangian boundary meshes. These boundaries interact via their coupling to the fluid flow, and hence one would be able to simulate confinement of these droplets without necessarily sacrificing the periodic solution method outlined in the Methods section. { A preliminary investigation into such external boundaries is discussed in \cite{Whitfield2015}. 

Finally, the modular nature of the code means that these simulations are capable of simulating a coupled active interface and an internal active polar fluid. The dynamics of such a system will be even more complex and so the results presented here will be relevant to understanding those systems in future. Similarly, this will be relevant also to systems of passive liquid crystal drops propelled by active surfactants as studied in \cite{Herminghaus2014}.}

In conclusion, the results we present here demonstrate complex dynamics in confined active fluids with minimal degrees of freedom. These simple one-component models demonstrate steady state dynamics akin to those observed in cells, cell fragments and reconstituted cell components that are driven by the active cell cytoskeleton. The numerical method we outline in this paper, based on the Immersed Boundary method enables continuum level simulation of these systems and has a wide range of potential applications in this field. 

\section*{Methods}
\label{sec:numerics}

In this section we outline explicitly the numerical method used in the simulations and the parameter value ranges studied.

\subsection*{Coordinate System}
\label{sec:coords}

\begin{figure}[h]
	\centering
	\includegraphics[width=0.6\textwidth]{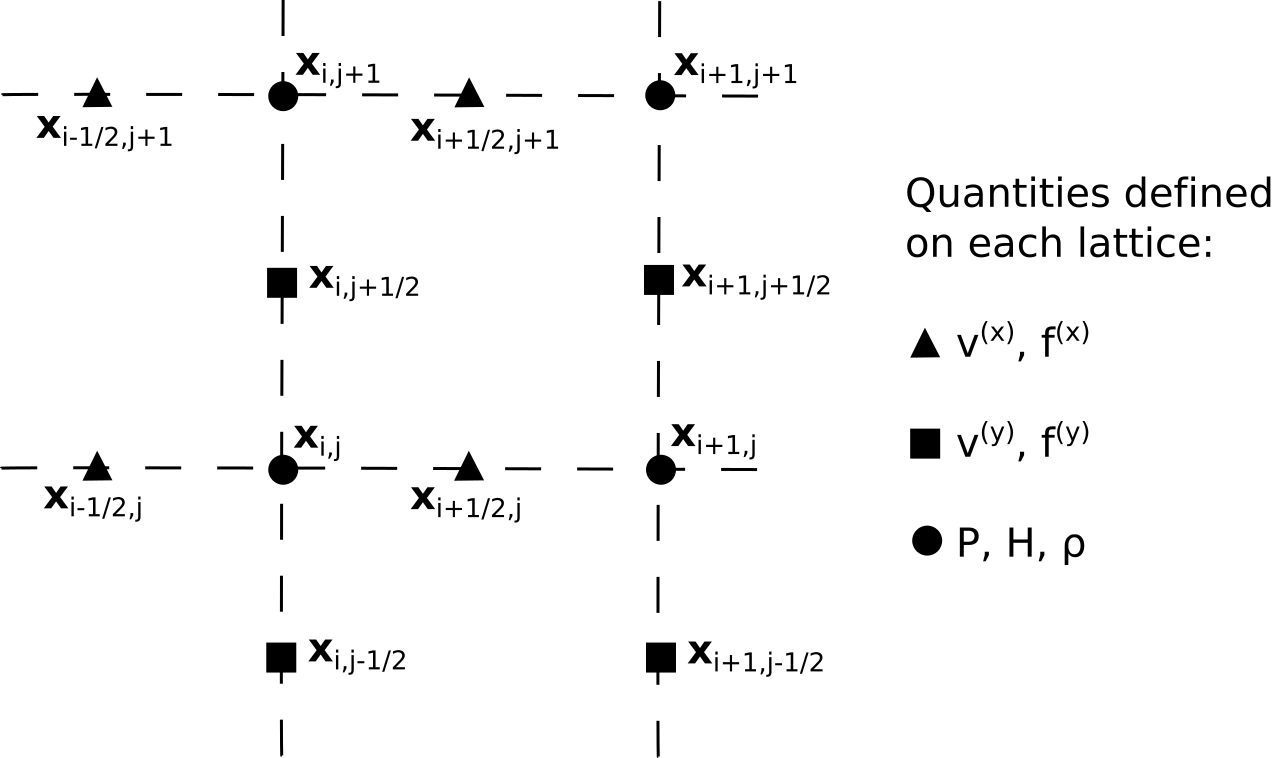}
	\centering
	\caption{Schematic of the staggered Cartesian grid used to model the fluid domain $\Omega$. Generally, $x$-components of vectors are on the $x_{i-1/2,j}$ grid,  $y$-components of vectors are on the $x_{i,j-1/2}$ grid and scalars are on the $x_{i,j}$ grid.}
    \label{fig:grid}
\end{figure}

The bulk fluid domain $\Omega$ is defined on a staggered periodic Cartesian grid of points with spacing $h$ in each direction. The 3 overlaid grids correspond to positions $(x_i,y_j)$, $(x_{i-1/2},y_{j})$, and $(x_{i},y_{j-1/2})$ where $x_{i-1/2} = x_i - h/2$ and the same for $y$. The points $x_i$ and $y_j$ run from $x_0 = -(L_x-h)/2$ and $y_0 = -(L_{y}-h)/2$ to $x_{M_x-1} = (L_x-h)/2$ and $y_{M_y-1} = (L_{y}-h)/2 - h$ (where $L_{x,y}$ are the lengths of the simulation box in the $x$ and $y$ directions and $M_{x,y} = L_{x,y}/h$ is the total number of grid points in each direction). The interface $C$ is described by a staggered grid of Lagrangian points $s_k$ where $s_k = k\Delta s$ with $k=0 \ldots N(t)-1$. These points have corresponding Cartesian coordinates $(X_k,Y_k)$ and $(X_{k+1/2},Y_{k+1/2}) = (X_{k+1}+X_{k},Y_{k+1}+Y_k)/2$. The step size is { defined} as $\Delta s = l_0/N(0)$ where $l_0$ is the initial boundary length and $N(0)$ the initial number of boundary points. Points can be added or removed from this mesh during the simulation, which will be outlined later in this section. Fig \ref{fig:grid} gives an outline of which quantities are defined on which grid. We integrate over time with step $\Delta t$ such that $t = n \Delta t$ for $n=0 \ldots T$. We will denote the time-step at which a quantity is defined by a superscript where necessary (\emph{e.g.} $\vect{X}^{n+1}_k$), if none is given then all quantities in that equation will be at the same arbitrary time-step $n$.

Staggered grids are commonly used for fluid simulations of this nature in order to achieve consistency between first and second order gradients without the need to use a 5-point stencil in 1D (or 13-point in 2D). We define the gradient operator on the bulk fluid as $\nabla_h g_{i,j} \equiv \rndb{\partial_x,\partial_y}g_{i,j} = (g_{i+1,j}-g_{i,j}\, , \,g_{i,j+1}-g_{i,j})/h$ so that if $\vect{G}_{i,j} = \nabla_h g_{i,j}$ then the $x$-component of the function $\vect{G}_{i,j}$ is defined on the grid $(x_{i+1/2}\, ,\, y_{j})$ and the $y$-component on $(x_i\, ,\, y_{j+1/2})$. Values of $\vect{G}_{i,j}$ can then be interpolated to other lattices where required. The Laplacian operator can then be defined as

\begin{align}
\label{gradient} \nabla^2_h g_{i,j} \equiv \nabla_h \cdot \rndb{\partial_x g_{i-1,j}\,,\,\partial_y g_{i,j-1}} = \frac{g_{i+1,j} + g_{i,j+1} - 4g_{i,j} + g_{i-1,j} + g_{i,j-1}}{h^2}
\end{align}
such that it is on the same grid as the function $g$. Similarly, the boundary derivative  $\partial_s g_{k} = (g_{k+1} - g_k)/\Delta s$ is defined on the $k+1/2$ lattice, such that $\partial_s^2 g_{k} = \partial_s (\partial_s g_{k-1}) = (g_{k+1}-2g_{k}+g_{k-1})\Delta s^2$.

\subsection*{Immersed Boundary Equations}

We use the 2D numerical delta function $\delta_h(\vect{r}-\vect{X})$ described in \cite{Peskin2002} to couple boundary quantities to the bulk fluid and vice versa. This delta function is given by the product of two 1-dimensional functions $\Phi$:

\begin{align}
	\delta_h(\vect{x}_{i,j}-\vect{X}_{k}) = \frac{\Phi\rndb{\sqb{x_i-X_k}/h}\Phi\rndb{\sqb{y_j-Y_k}/h}}{h^2}
\end{align}
where the function $\Phi$ acts as a numerical analogue of the 1D analytical Dirac delta function. It is desirable to use a function that is only non-zero for a few points, as this aids computational efficiency, hence a restriction regularly chosen is  $\Phi(r)=0$ if $r<-2$ or $r>2$. In \cite{Peskin2002} several constraints are imposed that one would expect from a delta function, namely that it be a continuous function in $r$ and:

\begin{align}
\label{phiprop1} \sum_{i = {\rm even}} \Phi(r-i) = \sum_{i = {\rm odd}} \Phi(r-i) &= \frac{1}{2} \\
\label{phiprop2} \sum_i (r-i)\Phi(r-i) &= 0 \\
\label{phiprop3} \sum_i \sqb{\Phi(r-i)}^2 &= \lambda
\end{align}
where $\lambda$ is a constant. Note that the condition in Eq \eqref{phiprop1} gives the identity $\sum_{i} \Phi(r-i) = 1$ but also ensures a symmetric distribution of $\Phi$. The derivation of $\Phi$ is outlined in reference \cite{Peskin2002}, here we just give the resulting function:

\begin{align}
\Phi(r) = 
\begin{cases}
\rndb{5 + 2r - \sqrt{-4r^2 - 12r - 7}}/8 \qquad &{\rm if} \quad {\rm -2} < r < {\rm -1}  \\
\rndb{3 + 2r + \sqrt{-4r^2-4r+1}}/8 \qquad &{\rm if} \quad {\rm -1} \leq r < 0 \\
\rndb{3 - 2r + \sqrt{-4r^2+4r+1}}/8 \qquad &{\rm if} \quad 0 \leq r < 1\\
\rndb{5 - 2r - \sqrt{-4r^2 + 12r - 7}}/8 \qquad &{\rm if} \quad 1 \leq r < 2\\
0 \qquad {\rm otherwise}
\end{cases} \, .
\end{align}
From this it is straightforward to check that $\lambda = 3/8$. It should be noted that this distribution maintains its properties for all real values of $r$ and is not restricted to integer values.\\

Using this delta function one can distribute the force density on the boundary $F^{(b)}_k$ to the fluid. First, we discretise Eq \eqref{Fb2D} to give:

\begin{align}
\label{Fbnum} F^{(b)}_{k+1} = \partial_s (\gamma_{k} \vect{\tau}_{k})\, ,
\end{align}
where the tangent vector (defined on the $\vect{X}_{k+1/2}$ grid) is given by
\begin{align}
\label{tang} \vect{\tau}_{k} = \frac{\partial_s \vect{X}_{k}}{\norm{\partial_s \vect{X}_{k}}} \, .
\end{align}
The corresponding force density components in the fluid then can be defined as:

\begin{align} 
\label{fx} f_{i,j}^{(x,b)} = \sum_{k=0}^{N-1} F_k^{(x,b)} \delta(\vect{x}_{i-1/2,j}-\vect{X}_k) \Delta s \\
\label{fy}  f_{i,j}^{(y,b)} = \sum_{k=0}^{N-1} F_k^{(y,b)} \delta(\vect{x}_{i,j-1/2}-\vect{X}_k) \Delta s \, ,
\end{align}
where the $(x)$ and $(y)$ superscripts are used to denote components of the vector. Note that the $x$ component is defined on the $\vect{x}_{i-1/2,j}$ grid and the $y$ component on the  $\vect{x}_{i,j-1/2}$, this is the same for the components of the velocity vector. We compute the velocity components $v^{(x)}_{i,j}$ and $v^{(y)}_{i,j}$, and pressure $P_{i,j}$ by solving the incompressible Stokes equation (i.e. Eq \eqref{fbfluid} such that $\divv \vect{v} = 0$). To do this we employ a projection method where we first define the intermediate solutions $\vect{v}'$ and $P'$ such that

\begin{align}
\label{veq} \nabla^2 \vect{v}' &= -\frac{1}{\eta}\vect{f}^{\rm tot} \\
\label{Peq} \nabla^2 P' &=  (\divv \vect{v}') \, ,
\end{align}
where $\vect{f}^{\rm tot}$ is the sum of all the force densities acting on the fluid. First, we solve for $\vect{v}'$; utilising the periodicity of the lattice we can write any function $g_{i,j}$ in terms of a 2D Discrete Fourier Transform (DFT):

\begin{align}
\label{DFT} g_{i,j} &= \frac{1}{\sqrt{M_x M_y}}\sum_{\alpha=0}^{M_x-1} \sum_{\beta=0}^{M_y-1} \hat{g}_{\alpha,\beta} \exp\sqb{2\pi I\rndb{\frac{i\alpha}{M_x} + \frac{j\beta}{M_y}}} \\
\label{DFT2} \hat{g}_{\alpha,\beta} &= \frac{1}{\sqrt{M_x M_y}}\sum_{i=0}^{M_x-1} \sum_{j=0}^{M_y-1} g_{i,j} \exp\sqb{-2\pi I\rndb{\frac{i\alpha}{M_x} + \frac{j\beta}{M_y}}} \,.
\end{align}
Here we have used the notation $I=\sqrt{-1}$ to avoid confusion with the index notation. Using Eqs \eqref{DFT} and \eqref{gradient} we can re-write Eqs \eqref{veq} and \eqref{Peq} so that for each value of $\alpha$ and $\beta$:

\begin{align}
\label{vxeqFT}  \hat{v'}^{(x)}_{\alpha,\beta}  &= \frac{1}{\eta \hat{D}^2_{\alpha,\beta}} \hat{f}^{(x,b)}_{\alpha,\beta} \\
\label{vyeqFT} \hat{v'}^{(y)}_{\alpha,\beta}  &= \frac{1}{\eta \hat{D}^2_{\alpha,\beta}} \hat{f}^{(y,b)}_{\alpha,\beta} \\
\label{PeqFT} \hat{P'}_{\alpha,\beta}  &= \frac{1}{\hat{D}^2_{\alpha,\beta}} \widehat{(\nabla_h \cdot \vect{v}')}_{\alpha,\beta}
\end{align}
where

\begin{align}
\label{Dop} \hat{D}^2_{\alpha,\beta} = \rndb{e^{\frac{2\pi I\alpha}{M_x}} + e^{\frac{2\pi I\beta}{M_y}} - 4 + e^{-\frac{2\pi I\alpha}{M_x}} + e^{-\frac{2\pi I\beta}{M_y}}} \, .
\end{align}
Using standard NAG library functions \cite{NAG} to compute the Fast Fourier Transforms (FFTs), we find the DFTs of the components of $\vect{f}^{\rm tot}_{i,j}$. Looping over all values of $\alpha$ and $\beta$ we can use Eqs \eqref{vxeqFT} and \eqref{vyeqFT} to assign values to the DFT of $\vect{v}'_{i,j}$. We then compute the inverse DFT of $\hat{\vect{v}}'$ to acquire values for $\vect{v}'_{i,j}$. Following this we calculate $\nabla_h \cdot \vect{v}'_{i,j}$ and compute its DFT. Then we can calculate $P'_{i,j}$ similarly. Finally, we relate these quantities back to the velocity and pressure as follows:

\begin{align}
\label{v_final}  \vect{v}_{i,j} &= \vect{v}'_{i,j} + \frac{1}{h}\rndb{P'_{i,j}-P'_{i-1,j},P'_{i,j}-P'_{i,j-1}}\\
\label{Pfinal} P_{i,j} &= \eta\nabla_h^2 P'_{i,j} \,.
\end{align}
This is a computationally efficient method for solving the Stokes equation as each set of assignments only require $M_x M_y$ operations. The time taken for the FFT computations scales approximately as $M_x M_y \log_{10}(M_x M_y)$ so this is the limiting complexity in this process.\\

Once the bulk velocity is computed we can calculate the velocity of the interface using Eq \eqref{vint}:

\begin{align}
	\label{Vbx} V_k^{(x)} = \sum_{i=i'-1}^{i'+2} \sum_{j=j'-1}^{j'+2} v_{i,j}^{(x)} \delta(\vect{x}_{i-1/2,j}-\vect{X}_k) h^2 \\
	\label{Vby} V_k^{(y)} = \sum_{i=i'-1}^{i'+2} \sum_{j=j'-1}^{j'+2} v_{i,j}^{(y)} \delta(\vect{x}_{i,j-1/2}-\vect{X}_k) h^2
\end{align}
where $i'$ is the value of $(X_k+(L_x-h)/2)/h$ rounded down to the nearest integer and $j'$ is the same but for $Y_k$. Generally in IB methods this interface is updated using a simple forward first order Euler scheme, \emph{i.e.} $\vect{X}^{n+1}_k = \vect{X}^n_k + \Delta t \vect{V}_k$. Instead we use an explicit fourth order Runge-Kutta method as outlined later in this section. We find that this method improves the numerical stability of the algorithm compared to a forward Euler update, particularly in the case where the boundary position is coupled to an internal polarisation field.

\subsection*{Active boundary equations}
\label{sec:ABeqns}

We define a concentration of active particles $c_{k}$ on the boundary points $\vect{X}_{k+1/2}$ which alters the surface tension $\gamma_{k}$ according to Eq \eqref{ST}. The concentration dynamics are then governed by Eq \eqref{ad_diff} which we discretise using the Crank-Nicholson scheme (as in \cite{Lai2008,Chen2014}):

\begin{align}
c_{k}^{n+1} \norm{\partial_s \vect{X}_{k}^{n+1}} =& \,c_{k}^{n} \norm{\partial_s \vect{X}_{k}^{n}}   + \frac{D\Delta t}{\Delta s}\rndb{\frac{\partial_s c_{k}^{n+1}}{\norm{\partial_s \vect{X}_{k+1}^{n+1}} + \norm{\partial_s \vect{X}_{k}^{n+1}}} - \frac{\partial_s c_{k-1}^{n+1}}{\norm{\partial_s \vect{X}_{k}^{n+1}} +  \norm{\partial_s \vect{X}_{k-1}^{n+1}}}} \notag \\
\label{conc} &+ \frac{D\Delta t}{\Delta s}\rndb{\frac{\partial_s c_{k}^{n}}{\norm{\partial_s \vect{X}_{k+1}^{n}} + \norm{\partial_s \vect{X}_{k}^{n}}} - \frac{\partial_s c_{k-1}^{n}}{\norm{\partial_s \vect{X}_{k}^{n}} +  \norm{\partial_s \vect{X}_{k-1}^{n}}}} + \frac{1}{2}\rndb{q_k^{n+1} + q_k^n}\Delta t \, .
\end{align}
The quantity $q$ contains the binding and unbinding rates for the concentration { and is defined further below in Eqs. \eqref{qkn} and \eqref{qkn1}}. We update the concentration $c$ after updating the boundary position $\vect{X}$ so that $\vect{X}^{n+1}$ is defined. Then the set of Eqs in \eqref{conc} can be represented by a periodic tridiagonal matrix:

\begin{align}
\label{tridiag} 
\begin{pmatrix}
d_{0} & u_{0} &  0 &  0 & l_{0} \\
l_{1} & d_{1} & u_{1} & 0 & 0  \\
0 &  \ddots & \ddots & \ddots & 0 \\
0  & 0 & l_{N-2} & d_{N-2} & u_{N-2} \\
u_{N-1} & 0 & 0 & l_{N-1} & d_{N-1} 
\end{pmatrix}
\begin{pmatrix}
c_{0}^{(n+1)} \\
c_{1}^{(n+1)} \\
\vdots \\
c_{N-2}^{(n+1)} \\
c_{N-1}^{(n+1)} \\
\end{pmatrix}
= 
\begin{pmatrix}
b_{0}^{(n+1)} \\
b_{1}^{(n+1)} \\
\vdots \\
b_{N-2}^{(n+1)} \\
b_{N-1}^{(n+1)} \\
\end{pmatrix} \,.
\end{align}
Rearranging Eq \eqref{conc} we can read off the expressions for $l$,$d$,$u$, and $b$:

\begin{align}
\notag d_{k} =& \rndb{1{+\frac{k_{\rm off}\Delta t}{2}}}\norm{\partial_s \vect{X}_{k+1}^{n+1}}\\ 
&+ \frac{D\Delta t}{(\Delta s)^2}\rndb{\frac{1}{\norm{\partial_s \vect{X}_{k+1}^{n+1}} + \norm{\partial_s \vect{X}_{k}^{n+1}}} - \frac{1}{\norm{\partial_s \vect{X}_{k}^{n+1}} +  \norm{\partial_s \vect{X}_{k-1}^{n+1}}}} \\
l_k =& -\frac{D\Delta t}{(\Delta s)^2 \rndb{\norm{\partial_s \vect{X}_{k+1}^{n+1}} + \norm{\partial_s \vect{X}_{k}^{n+1}}}}\\
u_k =& -\frac{D\Delta t}{(\Delta s)^2 \rndb{\norm{\partial_s \vect{X}_{k}^{n+1}} + \norm{\partial_s \vect{X}_{k-1}^{n+1}}}}\\
\notag b_k =& c_{k}^{n}\norm{\partial_s \vect{X}_{k}^{n}} + \frac{D\Delta t}{\Delta s}\rndb{\frac{\partial_s c_{k}^{n}}{\norm{\partial_s \vect{X}_{k+1}^{n}} + \norm{\partial_s \vect{X}_{k}^{n}}} - \frac{\partial_s c_{k-1}^{n}}{\norm{\partial_s \vect{X}_{k}^{n}} +  \norm{\partial_s \vect{X}_{k-1}^{n}}}}\\
& { + \frac{\Delta t}{2}(q_k^n + k_{\rm off}c_k^n + q_k^{n+1})}\, .
\end{align}
We solve this matrix directly using standard NAG linear algebra functions \cite{NAG}. We test that the mass is conserved numerically by outputting the change in total mass in the simulation between time-steps and find that these values are randomly distributed around zero with maximum deviations on the order of machine precision.

\subsection*{Coupling to a passive bulk fluid}
\label{sec:bulkconc} 

In order to numerically model the equations in the Model section we need a numerical analogue for the level-set function $H$. Taking the gradient of both sides of Eq \eqref{Hdefa} we see that by definition:

\begin{align}
\label{HPoisson} \nabla^2 H(\vect{x},t) = -\nabla \cdot \int_{C} \hat{\vect{n}} \delta(\vect{x}-\vect{X}){\rm d} l
\end{align}
Thus, as in  \cite{Unverdi1992,Chen2014}, to find the numerical equivalent of the function $H$ we solve the following Poisson equation:

\begin{align}
\label{Hcalc} \nabla_h^2 H_{i,j} = -\nabla_h \cdot \sum_{k=0}^{N(t)-1}  \sqb{\hat{n}^{(x)}_{k}\delta(\vect{x}_{i-1/2,j}-\vect{X}_{k+1/2}), \hat{n}^{(y)}_{k}\delta(\vect{x}_{i,j-1/2}-\vect{X}_{k+1/2})}\norm{\partial_s \vect{X}_k}\Delta s \, .
\end{align} 
We again solve this equation using the same FFT method used in to solve Eqs \eqref{vxeqFT}, \eqref{vyeqFT} and \eqref{PeqFT}. However, as we set the magnitude of the zero-wavenumber mode of the Fourier Transform to zero, this only defines $H$ up to a constant (more exactly it defines $H$ so that its average value is zero). This can be corrected simply in the simulation by recording the value of $H$ at any point where $\nabla H = 0$ and $H<0$ (a point strictly outside the drop) and then subtracting the value of $H$ at this point from the value of $H$ at all points in the simulation. This then results in a profile where all points outside the droplet are approximately equal to zero. However, as discussed in \cite{Teigen2009a} this can lead to numerical issues when dividing by $H$, and so we redefine the level set function at all points as $H=\sqrt{H^2+\epsilon^2}$ setting $\epsilon=10^{-6}$ as a sufficiently small discrepancy to avoid noticeable `leaking' of the internal concentration.

We can then define a concentration field that is contained within the droplet by the quantity $H\rho$. Thus, in order to conserve mass and ensure that the concentration is contained within the droplet we can use the same equations as used in \cite{Chen2014} to describe a soluble surfactant. We update $\rho$ from Eq \eqref{dcfdt} using the following Crank-Nicholson scheme (as used in \cite{Chen2014}):

\begin{align}
 \frac{1}{\Delta t}\sqb{(H\rho)^{n+1}_{i,j} - (H\rho)^{n}_{i,j}}  &= - \frac{1}{2}\vect{D}_h \cdot \sqb{(\vect{v}H\rho)^{n+1}_{i,j} + (\vect{v}H\rho)^{n}_{i,j}} \notag\\
&+ \frac{D_b}{2} \sqb{\vect{D}_h\cdot\rndb{H\vect{D}_h\rho}^{n+1}_{i,j} + \vect{D}_h\cdot\rndb{H\vect{D}_h\rho}^{n}_{i,j}} \notag \\
\label{drhodts} & - \frac{1}{2}\rndb{\sum_{k=0}^{N-1}q_k^{n+1}\delta_h(\vect{x}_{i,j} - \vect{X}_k^{n+1}) + \sum_{k=0}^{N-1}q_k^{n}\delta_h(\vect{x}_{i,j} - \vect{X}_k^{n})}
\end{align}
where $\vect{D}_h$ is a central difference operator on the staggered grid (as defined in \cite{Chen2014}). The exchange terms $q_k^n$ and $q_k^{n+1}$ are given by:

\begin{align}
\label{qkn} q_k^n &= k_{\rm on} \sum_{i=0}^{M_x}\sum_{j=0}^{M_y}H{i,j}^n \rho_{i,j}^n \delta_h(\vect{x}_{i,j} - \vect{X}_{k+1/2}^{n})h^2 - k_{\rm off}c_k^n \\
\label{qkn1} q_k^{n+1} &=  k_{\rm on} \sum_{i=0}^{M_x}\sum_{j=0}^{M_y}H{i,j}^{n+1} \rho_{i,j}^n \delta_h(\vect{x}_{i,j} - \vect{X}_{k+1/2}^{n+1})h^2 - k_{\rm off}c_k^{n+1} \, .
\end{align}
As we update the boundary positions and concentration before this step, $\vect{X}_k^{n+1}$, $c_k^{n+1}$ and $H_{i,j}^{n+1}$ are already defined.

Analytically we know that the system should reach a stable steady state where $\rho = k_{\rm off}c_0/k_{\rm on}$ inside the drop and $c=c_0$ on the interface. In these simulations however, the bulk concentration evaluated at the interface is $\sum_{i=0}^{M_x-1}\sum_{j=0}^{M_y-1}H_{i,j}^n \rho_{i,j}^n \delta_h(\vect{x}_{i,j} - \vect{X}_{k+1/2}^{n})h^2$, and so the equivalent steady state for this numerical system is given by:
\begin{align}
\label{rhoss} \rho_0 = \frac{k_{\rm off}c_0}{k_{\rm on} \left\langle \sum_{i=0}^{M_x}\sum_{j=0}^{M_y} H_{i,j} \rho_{i,j} \delta_h(\vect{x}_{i,j} - \vect{X}_{k+1/2})h^2 \right\rangle}  \, .
\end{align}
Where the droplet shape and position is assumed fixed in this passive case and the angled brackets indicate the average value over the boundary such that:
\begin{align}
\label{average} \left\langle f_k \right \rangle =  \frac{\sum_{k=0}^{N-1} f_k \norm{\partial \vect{X}/\partial s}\Delta s}{\sum_{k=0}^{N-1} \norm{\partial \vect{X}/\partial s}\Delta s} \, .
\end{align}
Thus, we initialise the simulation with a concentration $c_k=c_0=1$ on the interface and $\rho=0$ everywhere except at $x_i = y_j = 0$ where $\rho = \rho_0$ everywhere so that the concentration in the system is directly proportional to $H$ and is in the stable steady state configuration (in the absence of activity or external flows).

\subsection*{Active Polar Fluid}
\label{sec:afeqns}

We define the polarisation field $\vect{p}_{i,j}$, $\vect{h}_{i,j}$ and stress components $\tens{\sigma}^{(\rm act,dis,int)}_{i,j}$ from Eqs \eqref{Hact} - \eqref{sint} on the centred lattice $(x_{i},y_{j})$ and interpolate any gradients that are evaluated on the other lattices therein. First, we calculate the molecular field and the free energy density using \eqref{FreeEH} from the identity $\vect{h} = -\delta F_H/\delta \vect{p}$:

\begin{align}
	\label{molfield}  \vect{h}_{i,j} &= \frac{\delta F}{\delta \vect{p}} K\nabla_h^2 \vect{p}_{i,j} - Kc_b\rndb{\vect{p}^2_{i,j}-H_{i,j}}\vect{p}  - W_1 \sqb{\norm{\nabla_{hc}H_{i,j}} + \vect{p}_{i,j}\cdot\rndb{\nabla_{hc}H_{i,j}}}\nabla_{hc} H_{i,j}\\
	\notag g^{\rm tot}_{i,j} &= \frac{K}{2}\sqb{\rndb{\nabla_{hc} p^{(x)}_{i,j}}^2 + \rndb{\nabla_{hc} p^{(y)}_{i,j}}^2} + \frac{K}{4}c_b \vect{p}^2_{i,j}\rndb{\vect{p}^2_{i,j} - 2H_{i,j}}\\ 
	\label{freetot} &+ \frac{W_1}{2}\sqb{\norm{\nabla_{hc}H_{i,j}} + \vect{p}_{i,j}\cdot\rndb{\nabla_{hc}H_{i,j}}}^2
\end{align}
where the central gradient operator $\nabla_{hc}$ is defined by $\nabla_{hc} g_{i,j} = \rndb{g_{i+1,j}-g_{i-1,j},g_{i,j+1} - g_{i,j-1}}/2h$. From these we calculate the components of the stress $\tens{\sigma}^a_{i,j} = \tens{\sigma}^{\rm act}_{i,j} + \tens{\sigma}^{dis}_{i,j} + \tens{\sigma}^{int}_{i,j}$, and then finally the force density contribution from these stresses:

\begin{align}
	\label{fax} f^{(a,x)} = \frac{\sigma^{(a,xx)}_{i,j} - \sigma^{(a,xx)}_{i-1,j}}{h} + \frac{\sigma^{(a,yx)}_{i,j+1} - \sigma^{(a,yx)}_{i,j-1} + \sigma^{(a,yx)}_{i-1,j+1} - \sigma^{(a,yx)}_{i-1,j-1}}{4h} \\
	\label{fay} f^{(a,y)} = \frac{\sigma^{(a,yy)}_{i,j} - \sigma^{(a,yy)}_{i,j-1}}{h} + \frac{\sigma^{(a,xy)}_{i+1,j} - \sigma^{(a,xy)}_{i-1,j} + \sigma^{(a,xy)}_{i+1,j-1} - \sigma^{(a,xy)}_{i-1,j-1}}{4h}\, .
\end{align}
As with the velocity field, the $x$-component of the force density is defined on the $\vect{x}_{i-1/2,j}$ lattice and the $y$-component on the $\vect{x}_{i,j-1/2}$ lattice.
This force density is then added to the boundary force density in Eq \eqref{veq} before calculation of the velocity. Once the velocity is computed, the polarisation update equation can be computed from Eq \eqref{dpdta} as:

\begin{align}
	\label{dpxdt} \pdiff{p^{(x)}_{i,j}}{t} &= -\rndb{\vect{v}_{i,j} \cdot \nabla_{dw}}p^{(x)}_{i,j} - \omega_{i,j} p^{(y)}_{i,j} - \nu\rndb{u^{(xx)}_{i,j}p^{(x)} + u^{(xy)}_{i,j}p^{(y)}} + \frac{1}{\Gamma}h^{(x)}_{i,j} \\
	\label{dpydt} \pdiff{p^{(y)}_{i,j}}{t} &= -\rndb{\vect{v}_{i,j} \cdot \nabla_{dw}}p^{(y)}_{i,j} + \omega_{i,j} p^{(x)}_{i,j} - \nu\rndb{u^{(yx)}_{i,j}p^{(x)} + u^{(yy)}_{i,j}p^{(y)}} + \frac{1}{\Gamma}h^{(y)}_{i,j} 
\end{align}
where

\begin{align}
	\vect{v}_{i,j} =& \rndb{v^{(x)}_{i-1/2,j} + v^{(x)}_{i+1/2,j}\,,\,v^{(y)}_{i,j-1/2} + v^{(y)}_{i,j-1/2}} \notag \\
	\omega_{i,j} =& \frac{1}{4}\sqb{\rndb{\partial_x v^{(y)}_{i,j-1/2} + \partial_x v^{(y)}_{i,j+1/2} + \partial_x v^{(y)}_{i-1,j-1/2} + \partial_x v^{(y)}_{i-1,j+1/2}} \right. \notag  \\
		& \left.- \rndb{\partial_y v^{(x)}_{i-1/2,j} + \partial_y v^{(x)}_{i+1/2,j} + \partial_y v^{(x)}_{i-1/2,j-1} + \partial_y v^{(x)}_{i+1/2,j-1}}} \notag \\
	u^{(xx)}_{i,j} =& \partial_x v^{(x)}_{i-1/2,j} \notag \\
	u_{i,j}^{(xy)} = u_{i,j}^{(yx)} =& \frac{1}{4}\sqb{\rndb{\partial_x v^{(y)}_{i,j-1/2} + \partial_x v^{(y)}_{i,j+1/2} + \partial_x v^{(y)}_{i-1,j-1/2} + \partial_x v^{(y)}_{i-1,j+1/2}} \right. \notag  \\
		& \left.+ \rndb{\partial_y v^{(x)}_{i-1/2,j} + \partial_y v^{(x)}_{i+1/2,j} + \partial_y v^{(x)}_{i-1/2,j-1} + \partial_y v^{(x)}_{i+1/2,j-1}}} \notag \\
	u_{i,j}^{(yy)} =& \partial_y v_{i,j-1/2}^{(y)} \notag \,.
\end{align}
The polarisation is updated using the coupled fourth order Runge-Kutta method at the same time as the boundary is updated using Eqs \eqref{Vbx} and \eqref{Vby}, detailed in the ``Algorithm Overview'' section.

\subsection*{Adding and Removing Boundary Points}
\label{sec:addrem}

\begin{figure}[h]
	\centering
        \includegraphics[width=0.7\textwidth]{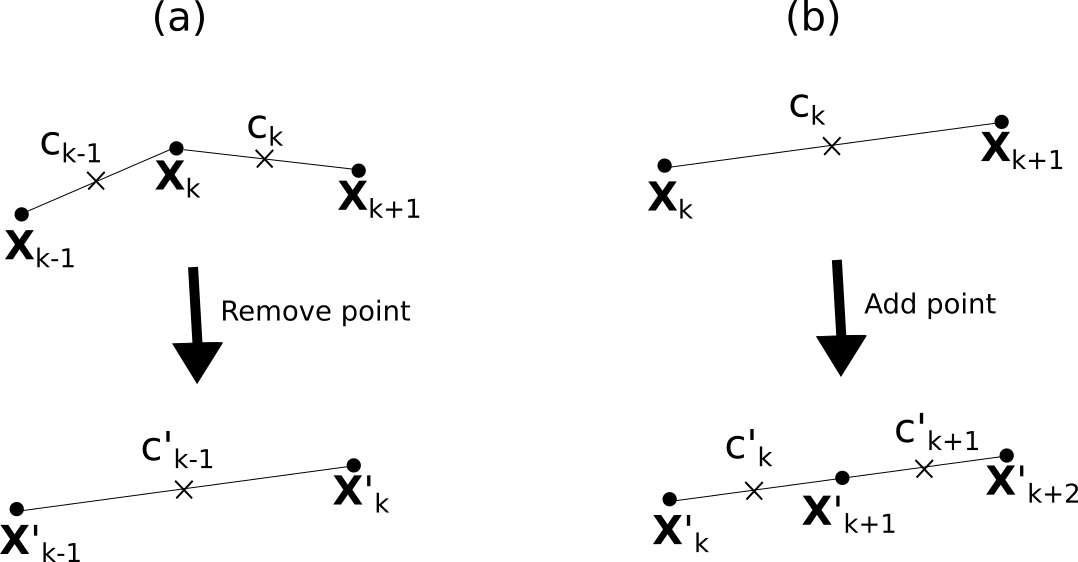}
        \caption{Schematic of \textbf{(a)} removing the point $\vect{X}_{k}$ and \textbf{(b)} adding a point between $\vect{X}_{k}$ and  $\vect{X}_{k+1}$. Circular points denote the boundary position nodes and crosses denote the concentration defined on the half-integer lattice.}
       \label{fig:addrem}
\end{figure}

As the immersed boundary in these simulations is the interface between two fluids, there is no resistance to neighbouring points getting too close or too far apart, as there would be in an elastic boundary. Therefore, it is necessary to add and remove points from the simulation to keep the boundary well defined and maintain accuracy. We remove the point $\vect{X}_{k}$ whenever $\norm{\partial_s \vect{X}_{k-1}} < 0.65$ and add a point between $\vect{X}_{k}$ and  $\vect{X}_{k+1}$ when $\norm{\partial_s \vect{X}_{k}} > 1.35$. See Fig \ref{fig:addrem} for a visual guide to this process.

To remove a point $\vect{X}_{k}$ we reassign $\vect{X}'_m = \vect{X}_{m+1}$ for all $k \le m < M-1$, then reduce the total number of boundary points to $N' = N-1$. Conservation of mass then dictates:

\begin{align}
\label{remove} c'_{k-1} = \frac{c_k \norm{\partial_s\vect{X}_k} + c_{k-1}\norm{\partial_s\vect{X}_{k-1}}}{\norm{\partial_s X'_{k-1}}}
\end{align}
and again $c'_{m} = c_{m+1}$ for all $k \le m < N-1$. The dashed quantities all denote the new values after removal of the boundary point.\\

To add a point between $\vect{X}_{k}$ and  $\vect{X}_{k+1}$ we reassign $\vect{X}'_{m+1} = \vect{X}_{m}$ for all $k<m<N$ then add the point $\vect{X}'_{k+1} = (\vect{X}_{k+1} + \vect{X}_k)/2$ with concentration $c'_{k+1} = c_k$ to conserve mass ($c_k$ remains unchanged).

We notice that when large flow fields are generated in the simulations, inaccuracies in the boundary positions can accumulate that need to be addressed. Firstly, we observe `buckling' of neighbouring boundary points in areas where the boundary flow converges, which results in points not being removed from the simulation even though they are overlapping. Thus, we also add a condition for the removal of a boundary point if $\norm{\nabla_s^2 R_k} > d_{\rm max}$ where, after some testing of different boundary shapes, we arrived at $d_{\rm max} = 0.05 R_0/ \Delta s^2$, where we have defined $R_k$ to be the distance of the boundary point from the droplet centre of mass. Therefore any excessive second order gradients in the boundary position are effectively smoothed out. However, we choose the limit so that smooth boundary deformations are still permitted, such as those observed in our results for an active polar fluid drop.\\
	
\subsection*{Droplet Area Conservation}
\label{sec:areacon}

We observe that the droplets can lose significant area over the whole simulation time, despite the fluid incompressibility. This effect is observed in the fluid immersed boundary simulations presented in \cite{Lai2008} and \cite{Chen2014} but is very slight and gradual (on the order of $10^{-4}\%$ area loss over a whole simulation) as the size of the flows is imposed externally. It is possible to observe much more rapid area loss in our simulations due to the continual addition and removal of interface points, and so to prevent this, and be consistent with the incompressibility condition, we calculate the area at each time step as

\begin{align}
\label{droparea} A^n =  \sum_{i=0}^{M_x-1}  \sum_{j=0}^{M_y-1} H^n_{i,j} h^2 \,.
\end{align}
We subtract this from the initial droplet area at time $n=0$ to get ${\rm d}A = A_0-A_n$. In order to keep the shape fixed, we then alter the values of $\vect{X}_k$ with a constant in the normal direction: $\delta \vect{X} = (\sqrt{(A^n-dA)/\pi} - \sqrt{(A^0/\pi)})\hat{\vect{n}}$. We then iterate this process, recalculating the area after this alteration and updating the value of $\delta \vect{X}$ accordingly until the area is conserved within a certain tolerance. We choose this tolerance to be $10^{-6}A^0$ and find that this step rarely takes more than 2 iterations, and in an individual step $\delta \vect{X}$ is at most the order of $10^{-6}R_0$, which is several orders of magnitude below the average velocity of the points in these simulations. Thus this area conservation step makes little difference to the evolution of the droplet.

\section*{Overview of the Algorithm}
\label{sec:overview}

The final simulation code is modular in nature, so that we can turn features on or off. The bare code contains the simulation of a passive fluid drop in a periodic domain, then one can choose whether to add an active boundary and an active polar fluid separately. On initialisation we define the fluid boundary points $\vect{X}_k$, the concentration on the boundary $c_{k+1/2}$ (if included) and the filament polarisation $\vect{p}_{i,j}$ (if included). The order of execution of the algorithm is given below:

\begin{enumerate}
	\item Boundary forces calculated according to Eq \eqref{Fbnum}.
	\item Bulk forces calculated as outlined in Eqs \eqref{molfield}-\eqref{fay} (if active polar fluid is included).
	\item Solve the Stokes' equation following Eqs \eqref{veq} to \eqref{Pfinal}.
	\item Calculate boundary velocity using Eqs \eqref{Vbx} and \eqref{Vby}.
	\item Simultaneously update boundary and polarisation field using an explicit fourth-order Runge-Kutta method (outlined below).
	\item Correct the boundary position for area conservation as outlined in the previous section.
	\item Update concentration at the interface by solving the system in Eq \eqref{tridiag}. 
	\item Update the bulk concentration $\rho$ using the Crank-Nicholson scheme of Eq \eqref{drhodts}.
	\item Add or remove fluid boundary points as outlined by Fig \ref{fig:addrem}.
\end{enumerate}
In step 5 we apply a coupled fourth order Runge-Kutta method where we update the boundary and the polarisation field. At the start of the update we store $\vect{p}_{i,j}^n$ and $\vect{X}_{i,j}^n$, then update them for a half time-step such that $\vect{p}_{i,j}^{n+1/4} = \vect{p}_{i,j}^n + \Delta t \vect{k}_1/2$ and  $\vect{X}_{k}^{n+1/4} = \vect{X}_{k}^n + \Delta t \vect{V}^n_{k}/2$ where $\vect{k}_1$ is the vector given by the right hand sides of Eqs \eqref{dpxdt} and \eqref{dpydt}. We then repeat steps 1-4 for this new polarisation and boundary position. Repeating this process we can define:

\begin{align}
	\label{k2} \vect{p}_{i,j}^{n+1/2} = \vect{p}_{i,j}^n + \frac{\Delta t}{2} \vect{k}_2 \qquad & \qquad \vect{X}_{k}^{n+1/2} = \vect{X}_{k}^n + \frac{\Delta t}{2} \vect{V}^{n+1/4}_{k}\\
	\label{k3} \vect{p}_{i,j}^{n+3/4} = \vect{p}_{i,j}^n + \Delta t \vect{k}_3 \qquad & \qquad \vect{X}_{k}^{n+3/4} = \vect{X}_{k}^n + \Delta t \vect{V}^{n+1/2}_{k}
\end{align}
where $\vect{k}_2$ and $\vect{k}_3$ are again the right hand sides of Eqs \eqref{dpxdt} and \eqref{dpydt} evaluated at the $n+1/4$ and $n+1/2$ time-steps respectively. We also define $\vect{k}_4$ similarly as this vector evaluated for the $n+3/4$ values of the polarisation and boundary condition. In between each fractional update we repeat steps 1-4. Then finally the updated polarisation and boundary are given by:

\begin{align}
	\label{pnew} \vect{p}_{i,j}^{n+1} &= \vect{p}_{i,j}^n + \frac{\Delta t}{6} \rndb{\vect{k}_1 + 2\vect{k}_2 + 2\vect{k}_3 + \vect{k}_4} \\
	\label{Xnew} \vect{X}_{k}^{n+1} &= \vect{X}_{k}^n + \frac{\Delta t}{6}\rndb{ \vect{V}^{n}_{k} + 2 \vect{V}^{n+1/4}_{k} + 2\vect{V}^{n+1/2}_{k} + \vect{V}^{n+3/4}_{k} }
\end{align}

We run the simulation, iterating over steps 1-9 usually $10^6$ times, unless we deem that longer is required. A full list of the parameter value ranges is given below:

\begin{table}[h]
	\centering
\begin{tabular}{|c|c|c|}
	\hline
	\textbf{Parameter name} & \textbf{Simulation values} & \textbf{SI units}\\
	\hline
	Fluid viscosity $\eta$ & 1 & 10 k Pa s\\
	\hline
	Passive surface tension $\gamma$ & 0.1 - 2.5 & 1 - 25 k Pa $\mu$m\\
	\hline
	Initial droplet radius $R_0$ & 1 & 10 $\mu$m\\
	\hline
\end{tabular}	
\caption{Fluid simulation parameters}
\label{Fluid params}
\end{table}	

\begin{table}[h]
	\centering
	\begin{tabular}{|c|c|c|}
		\hline
		\textbf{Parameter name} & \textbf{Simulation values} & \textbf{SI units}\\
		\hline
		Interface activity $\actc$ & $-2$ - 0 & -20 - 0 k Pa $\mu$m$^2$\\
		\hline
		Interface concentration $c^0$ & 1 & 0.1 $\mu$m$^{-1}$ \\
		\hline
		Interface diffusion coefficient $D$ & 0.1 - 0.5 & 1 - 5 $\mu$m$^2$s$^{-1}$\\
		\hline
		Interface passive pressure $B$ & 0 - 1 & 0 - 100 k Pa $\mu$m$^3$ \\
		\hline
		Bulk diffusion parameter $D_f$ & 0.001 - 0.5 & 0.01 - 5 $\mu$m$^2$s$^{-1}$ \\
		\hline
		Unbinding rate $k_{\rm off}$ & 0.01 - 1.0 & 10$^{-3}$ - 0.1 s$^{-1}$\\
		\hline 
		Binding rate $k_{\rm on}$ & 0.01 - 1.0 & 0.01 - 1 $\mu$m s$^{-1}$\\
		\hline
	\end{tabular}
	\caption{Active boundary parameters}
	\label{ABparams}
\end{table}
	
\begin{table}[h]
\centering
\begin{tabular}{|c|c|c|}
	\hline
	\textbf{Parameter name} & \textbf{Simulation values} & \textbf{SI units}\\
	\hline
	Rotational viscosity $\Gamma$ & 1.0 & 10 k Pa s \\	
	\hline
	Bulk activity $\act$ & $-10$ - 5 & $-10$ - 5 k Pa \\
	\hline
	Elastic constant $K$ & 0.01 - 0.1 & 1 - 10 nN\\
	\hline
	Anchoring constant $W_1$ & 0 and 1.25 &  0 - 12.5 nN $\mu$m$^{-1}$\\
	\hline
	Effective concentration $c_b$ & 50 & 0.05 $\mu$m$^{-2}$\\
	\hline
	Liquid crystal coupling constant $\nu$ & $\pm 1.1$ & dimensionless\\
	\hline
\end{tabular}
\caption{Active polar fluid parameters}
\label{APFparams}
\end{table}

\section*{Acknowledgements}

We acknowledge the EPSRC for funding this work, grant reference EP- K503149-1. We also thank Suzanne Fielding for useful discussion and insight, and Davide Marenduzzo and Elsen Tjhung for sharing the simulation source code used in \cite{Tjhung2012}.

\section{Supplementary Information}

\paragraph*{S1 Video.}
\label{S1_Video}
{\bf (a).}  Steady state swimming droplet from figure \ref{fig:bulkvel}(a). {\bf (b).} `Wandering' droplet from figure \ref{fig:bulkvel}(b).  {\bf (c).} Steady symmetric droplet from figure \ref{fig:bulkvel}(c). Colour gradients show concentration on the interface and in the bulk, from black (low) to yellow (high).

\paragraph*{S2 Video.}
\label{S2_Video}
Extensile active polar droplet with strong anchoring transitions to steady motile state. Corresponds to figure \ref{fig:shapevel}.

\paragraph*{S3 Video.}
\label{S3_Video}
Contractile active polar droplet with strong anchoring transitions to symmetric deformed configuration. Corresponds to figure \ref{fig:deformed}.

\paragraph*{S4 Video.}
\label{S4_Video}
Contractile active polar droplet with low Frank Elasticity ($K=10^{-3}$) displays active turbulent motion. Corresponds to figure \ref{fig:turbulence}.

\end{document}